\documentclass[fleqn,usenatbib]{mnras}

\usepackage{mathptmx}

\usepackage[T1]{fontenc}
\usepackage{ae,aecompl}

\usepackage{graphicx}	
\usepackage{amsmath}	
\usepackage{amssymb}	
\usepackage{colortbl}    
\usepackage{mathrsfs}

  \title[Variability of Onfp/Oef stars]{Similar but different: the varied landscape of Onfp/Oef stars variability\thanks{Based on data obtained with the {\it TESS} mission, whose funding is provided by the NASA Explorer Program, and on spectra collected at the Observatoire de Haute Provence.}}

  \author[G.\,Rauw \& Y.\,Naz\'e]{Gregor Rauw$^1$\thanks{E-mail: g.rauw@uliege.be}, Ya\"el Naz\'e$^1$\thanks{Senior Research Associate FRS-FNRS (Belgium)}
    \\
$^1$Space sciences, Technologies and Astrophysics Research (STAR) Institute, Universit\'e de Li\`ege, All\'ee du 6 Ao\^ut, 19c, B\^at B5c,\\ 4000 Li\`ege, Belgium}

\date{Accepted XXX. Received YYY; in original form ZZZ}

\pubyear{2020}

\begin{document}
\label{firstpage}
\pagerange{\pageref{firstpage}--\pageref{lastpage}}
\maketitle
\begin{abstract}
  The Oef category gathers rapidly rotating and evolved O-stars displaying a centrally reversed He\,{\sc ii} $\lambda$~4686 emission line. The origin of the variability of their photospheric and wind spectral lines is debated, with rotational modulation or pulsations as the main contenders. To shed new light on this question, we analysed high-quality and high-cadence {\it TESS} photometric time series for five Oef stars. We also collected a new time series of spectra for one target ($\lambda$~Cep) which had been the subject of specific debates in the last years. These observations reveal the variety of Oef behaviours. While space-based photometric data reveal substantial red noise components in all targets, only $\zeta$~Pup seems to display a long-lived periodicity. In our sample, stars exhibit a dominant signal at low frequencies but it appears relatively short-lived. This is reminiscent of rotational modulations by transient photospheric spots, though this scenario is challenged by the case of HD~14\,442, whose 1.230\,d$^{-1}$ signal significantly exceeds the critical rotational frequency. In parallel, no evidence of persistent $p$-mode non-radial pulsations is found in either photometry or spectroscopy of the stars, only temporary excitation of $g$-mode pulsations could offer an alternative explanation for the dominant signals. Finally, the revised luminosities of the stars using {\it GAIA-DR2} show that they are not all supergiants as $\zeta$~Pup. The question then arises whether the Oef peculiarity denotes a homogeneous class of objects after all.
\end{abstract}


\begin{table*}
  \caption{Key properties of our target stars}
  \begin{tabular}{c c r c c c c}
    \hline
    Star & $m_V$ & \multicolumn{1}{c}{$B-V$} & \multicolumn{3}{c}{Spectral type} & $\varpi$ \\
    \cline{4-6}
    &  &  &  Conti & Walborn & GOSC & (mas) \\
    \hline\hline
    $\lambda$~Cep & $5.04 \pm 0.02$ & $0.25 \pm 0.01$ & O6.5\,Ief & O6\,I(n)fp & O6.5\,I(n)fp  & $1.620 \pm 0.127$ \\
    HD~14\,434   & $8.50 \pm 0.01$ & $0.16 \pm 0.01$ & O6.5(ef) & O5.5\,Vn((f))p& O5.5\,IVnn(f)p & $0.391 \pm 0.047$ \\
    HD~14\,442   & $9.22 \pm 0.02$ & $0.40 \pm 0.02$ & O5.5ef & O5n(f)p & O5n(f)p & $0.256 \pm 0.034$ \\
    HD~192\,281   & $7.55 \pm 0.01$ & $0.37 \pm 0.02$ & O5(ef) & O5\,Vn((f))p & O4.5\,IV(n)(f) & $0.751 \pm 0.033$ \\
    BD$+60^{\circ}$\,2522 & $8.66 \pm 0.03$ & $0.40 \pm 0.03$ & O6.5\,IIIef & O6.5(n)(f)p & O6.5(n)fp & $0.371 \pm 0.029$ \\
    \hline
    HD~93\,521   & $7.03 \pm 0.01$ & $-0.28 \pm 0.01$ & O9.5\,Vp & O9.5\,IIInn & O9.5\,IIInn & $0.513 \pm 0.123$ \\
    \hline
  \end{tabular}
  \label{keyprop}
  
  {\scriptsize Apparent $V$ magnitude and $B-V$ colour are taken from the compilation of \citet{Reed}. For the Oef stars, the `Conti' spectral types are taken from \citet{Rau15}, \citet{DeB04} and \citet{Rau03}, whilst the `Walborn' spectral types come from \citet{Wal10}, and the General O-Star Catalogue (GOSC) classifications are taken from \citet{Holgado}. The trigonometric parallaxes come from the second {\it GAIA} data release \citep{Brown}.}  
\end {table*}
\begin{table*}
  \caption{Spectroscopic parameters of our target stars}
  \begin{tabular}{c c c c c c c c c c}
    \hline
    Star & \multicolumn{4}{c}{Cazorla et al.\ 2017} & & \multicolumn{4}{c}{Holgado et al.\ 2020} \\
    \cline{2-5}  \cline{7-10}
         & T$_{\rm eff}$ & $\log{g}$ & $\log{g_C}$ & $v\,\sin{i}$ & & T$_{\rm eff}$ & $\log{g}$ & $\log{g_C}$ & $v\,\sin{i}$ \\
         &  (kK) & (cgs) & (cgs) &  (km\,s$^{-1}$) & &  (kK) & (cgs) & (cgs) &  (km\,s$^{-1}$)  \\
    \hline\hline
    $\lambda$~Cep & $36.0 \pm 1.5$ & $3.50 \pm 0.15$ & $3.56 \pm 0.15$ & $214 \pm 15$ & & $35.8 \pm 0.5$ & $3.39 \pm 0.06$ & $3.47 \pm 0.04$ & $214$\\
    HD~14\,434   &  $40.0 \pm 1.5$ & $3.89 \pm 0.15$ & $4.03 \pm 0.15$ & $408 \pm 15$ & & $38.6 \pm 1.1$ & $3.78 \pm 0.17$ & $3.96 \pm 0.11$ & $417$ \\
    HD~14\,442   &  $39.2 \pm 1.5$ & $3.69 \pm 0.15$ & $3.78 \pm 0.15$ & $285 \pm 15$ & & $39.1 \pm 1.3$ & $3.62 \pm 0.13$ &                 & $324$ \\
    HD~192\,281   & $39.0 \pm 1.5$ & $3.64 \pm 0.15$ & $3.73 \pm 0.15$ & $276 \pm 15$ & & $40.8 \pm 1.1$ & $3.73 \pm 0.09$ & $3.82 \pm 0.07$ & $277$ \\
    BD$+60^{\circ}$\,2522 & $37.0 \pm 2.0^*$ & & & $239 \pm 11^*$ & & $36.2 \pm 1.1$ & $3.55 \pm 0.14$ &           & $247$ \\
    \hline
    HD~93\,521   &  $30.0 \pm 1.0$ & $3.60 \pm 0.10$ & $3.78 \pm 0.10$ & $405 \pm 15$ & & $31.7 \pm 0.8$ & $3.54 \pm 0.14$ & $3.78 \pm 0.08$ & $385$ \\
    \hline
  \end{tabular}
  \label{specprop}
  
  {\scriptsize The spectroscopic parameters in columns 2 -- 5 are taken from \citet{Cazorla}, except for BD$+60^{\circ}$\,2522 whose $v\,\sin{i}$ and effective temperature (listed with an asterisk) were taken from \citet{Pen09} and from the effective temperature of an O6.5\,III from \citet{Martins}, respectively. The parameters in columns 6 -- 9 are from \citet{Hol20} and \citet{Holgado}. Spectral analyses by \citet{Cazorla} were done with the CMFGEN model atmosphere code \citep{CMFGEN}, except for HD~93\,521 which was analysed with DETAIL/SURFACE \citep{Butler}. Spectral analyses by \citet{Holgado} and \citet{Hol20} were performed with the FASTWIND code \citep{FASTWIND}. The $g_C$ values indicate gravities corrected for the effects of the centrifugal forces.}  
\end {table*}
\begin{table*}
  \caption{Journal of the {\it TESS} observations of our target stars}
  \begin{tabular}{c c c c c c}
    \hline
    Star & Sector & Dates & Camera & Type & N \\
         &        & (BJD$-$2\,450\,000) & & & \\
    \hline\hline
    $\lambda$~Cep & 16 & 8\,738.6 -- 8\,763.3 & 2 & LC & 15\,057 \\
                  & 17 & 8\,764.7 -- 8\,788.5 & 3 & LC & 15\,169 \\
                  & 24 & 8\,955.8 -- 8\,982.3 & 4 & LC & 18\,061 \\
    \hline
    HD~14\,434    & 18 & 8\,790.7 -- 8\,814.1 & 2 & LC & 14\,872 \\
    \hline
    HD~14\,442    & 18 & 8\,790.7 -- 8\,814.6 & 2 & FFI & 1\,069 \\
    \hline
    HD~192\,281   & 14 & 8\,683.4 -- 8\,710.2 & 1 & LC & 13\,978 \\
                  & 15 & 8\,711.4 -- 8\,737.4 & 1 & LC & 13\,479 \\
    \hline
    BD$+60^{\circ}$\,2522 & 17 & 8\,764.7 -- 8\,788.7 & 3 & FFI & 1\,086 \\
                        & 24 & 8\,955.8 -- 8\,982.3 & 4 & FFI & 1\,185 \\
    \hline
    HD~93\,521    & 21 & 8\,870.5 -- 8\,897.8 & 1 & FFI & 1\,098 \\
    \hline
  \end{tabular}
  \label{journal}
  
  {\scriptsize In column 5, the LC and FFI labels stand respectively for 2\,min high-cadence lightcurves and full frame images collected every 30\,min. For the 2\,min lightcurves of sectors 14 -- 18, we used data release 30. The last column yields the number of good data points retained in the analysis.} 
\end {table*}
\section{Introduction}
Spectral line profile variability is a widespread property of single O-type stars \citep[e.g.][]{FGB}. Whilst photospheric lines can undergo such variations either as a result of non-radial pulsations \citep[NRPs, e.g.][]{Jan00,Rau08} or spots rotating with the star \citep[e.g.][]{SudHen16}, lines formed in the stellar wind can display variability as a result of moving small-scale clumps \citep[e.g.][]{Eversberg,Lep08} or large-scale structures, either magnetically confined winds \citep[e.g.][]{Sta96,Naz10} or corotating interaction regions (CIRs) most likely resulting from bright spots or NRPs \citep[e.g.][]{How93,Cran96,LB08,MP15,DaUr17}. 

Over recent years, access to space-borne high-precision photometry led to the detection of low-level photometric variability for a number of O-type stars \cite[e.g.][]{Bri11,Mahy,Blomme,2Ian,How14,Buy15,Aerts,Tahina,Bow19,Bow20}. Various causes of variability were found, acting simultaneously or not in a given star: binarity effects, rotational modulation, stochastic or multiperiodic ($\beta$~Cep-like) NRP pressure modes, internal gravity waves, etc.\ \citep[e.g.][]{Buy15,Bow19,Bow20,Burs}. 

Since rotational modulation can play an important role in these variability processes, rapidly rotating O-stars are of special interest. One of the most intensively studied objects in this context is the O4\,I(n)fp (or O4\,Ief) supergiant $\zeta$~Pup \citep[][and references therein]{2Ian,Tahina}. This star belongs to the scarce category of Onfp stars introduced by \citet{Wal73} and subsequently renamed Oef stars by \citet{CL74}. The main distinctive feature of these stars is their centrally reversed He\,{\sc ii} $\lambda$\,4686 emission line \citep{Wal73,CL74,Wal10} which is attributed to the effect of rotation \citep{Hil12}. Another common feature is the presence of significant spectroscopic variability, especially in this He\,{\sc ii} $\lambda$\,4686 line, e.g. $\lambda$~Cep \citep{Kap97,Rau15,SudHen16}, BD$+60^{\circ}$\,2522 \citep{Rau03}, or HD~192\,281, HD~14\,442 and HD~14\,434 \citep{DeB04}. Usually, the timescale as well as the pattern of the spectral variability strongly change from one epoch to the other \citep{Kap99,Rau03,SudHen16}.

The best studied object of this class is $\zeta$~Pup, whose photometric variability properties are also well constrained now. Indeed, space-borne photometry allowed the discovery \citep{2Ian} and subsequent confirmation \citep{Tahina,Burs} of a periodicity near 1.78\,d. Whilst the pattern of the photometric modulation somewhat changes with time, the 1.78\,d period appears stable, as it was found in independent data collected more than a decade apart ({\it SMEI} 2003--2006, {\it BRITE} 2014--now, {\it TESS} 2019, though it could have been absent at the time of {\it Hipparcos} observations, \citealt{how19}). This 1.78\,d cycle was thus interpreted as the rotational period of a spotted star \citep{Tahina}, though pulsations remain as contender \citep{2Ian,how19}. In the present paper, we investigate the photometric variability of the other historical members of the Oef category in our Galaxy ($\lambda$~Cep, HD~14\,434, HD~14\,442, HD~192\,281 and BD$+60^{\circ}$\,2522, see Tables\,\ref{keyprop} and \ref{specprop}) by means of data collected with NASA's {\it TESS} satellite. Together with $\zeta$~Pup, our targets make-up the full sample of Galactic Oef stars defined by \citet{CL74} and for which variability can be studied\footnote{More recently, HD~172\,175 was found to be another member of the Onfp/Oef class \citep{Wal10,Sota11}, but since neither {\it TESS} data nor an investigation of spectrocopic variability exist for this star, we do not consider it any further.}. For comparison, we also present the {\it TESS} lightcurve of the rapidly rotating, non-radially pulsating O9.5\,Vp star HD~93\,521 \citep{How98,Rau08}.

\section{Observations and data processing} \label{obs}
\subsection{Photometry}
The Transiting Exoplanet Survey Satellite \citep[{\it TESS},][]{TESS} is an all-sky survey space telescope operated by NASA. The primary goal of the mission is the detection of transiting exoplanets over a very wide field of view. As a by-product, {\it TESS} provides high-precision photometry for a huge number of stars either every 2\,minutes for objects declared as of particular interest or every 30\,minutes through full-frame images of the entire field of view. {\it TESS} observes the sky in sectors measuring $24^{\circ} \times 96^{\circ}$. Each sector is observed for two consecutive spacecraft orbits, i.e.\ for about 27\,days. Each lightcurve has a gap in the middle of the sector corresponding to an interruption of the observations for data downlink around perigee passage.

The bandpass of the {\it TESS} instrument ranges from 6000\,\AA\, to 1\,$\mu$m \citep{TESS}. The $15 \times 15\,\mu$m$^2$ pixels of the CCD detectors (corresponding to (21\arcsec )$^2$ on the sky) undersample the instrument PSF but photometry is always extracted over several pixels. Furthermore, bright objects, such as $\lambda$~Cep ($m_V = 5.04$) saturate the central pixel. However, the excess charges are spread into adjacent pixels via the blooming effect. This effect is used in the pipeline processing to recover the photometry for stars up to fourth magnitude. For a rather isolated source such as $\lambda$~Cep, we expect this correction to work out well. The {\it TESS} pipeline \citep{Jen16} is based on the one designed for the {\it Kepler} mission \citep{Jen10}. 

The journal of the {\it TESS} observations of our targets is provided in Table\,\ref{journal}. The high-cadence photometric lightcurves of $\lambda$~Cep, HD~14\,434 and HD~192\,281 were retrieved from the Mikulski Archive for Space Telescopes (MAST) portal\footnote{http://mast.stsci.edu/}. These lightcurves provide simple background-corrected aperture photometry as well as so-called PDC photometry obtained after removal of trends that correlate with systematic spacecraft or instrument effects. We discarded all data points with a quality flag different from 0. The formal photometric accuracies of the PDC data are 0.08\,mmag, 0.26\,mmag and 0.48\,mmag for $\lambda$~Cep, HD~192\,281, and HD~14\,434, respectively. With a nominal time step of 2\,min, these time series have a Nyquist frequency of 360\,d$^{-1}$. For targets with observations over several sectors, the mean PDC magnitudes can differ between consecutive sectors. For our combined analysis, we then subtracted the mean magnitude from each sector.
\begin{figure}
\begin{center}
  \resizebox{8.5cm}{!}{\includegraphics{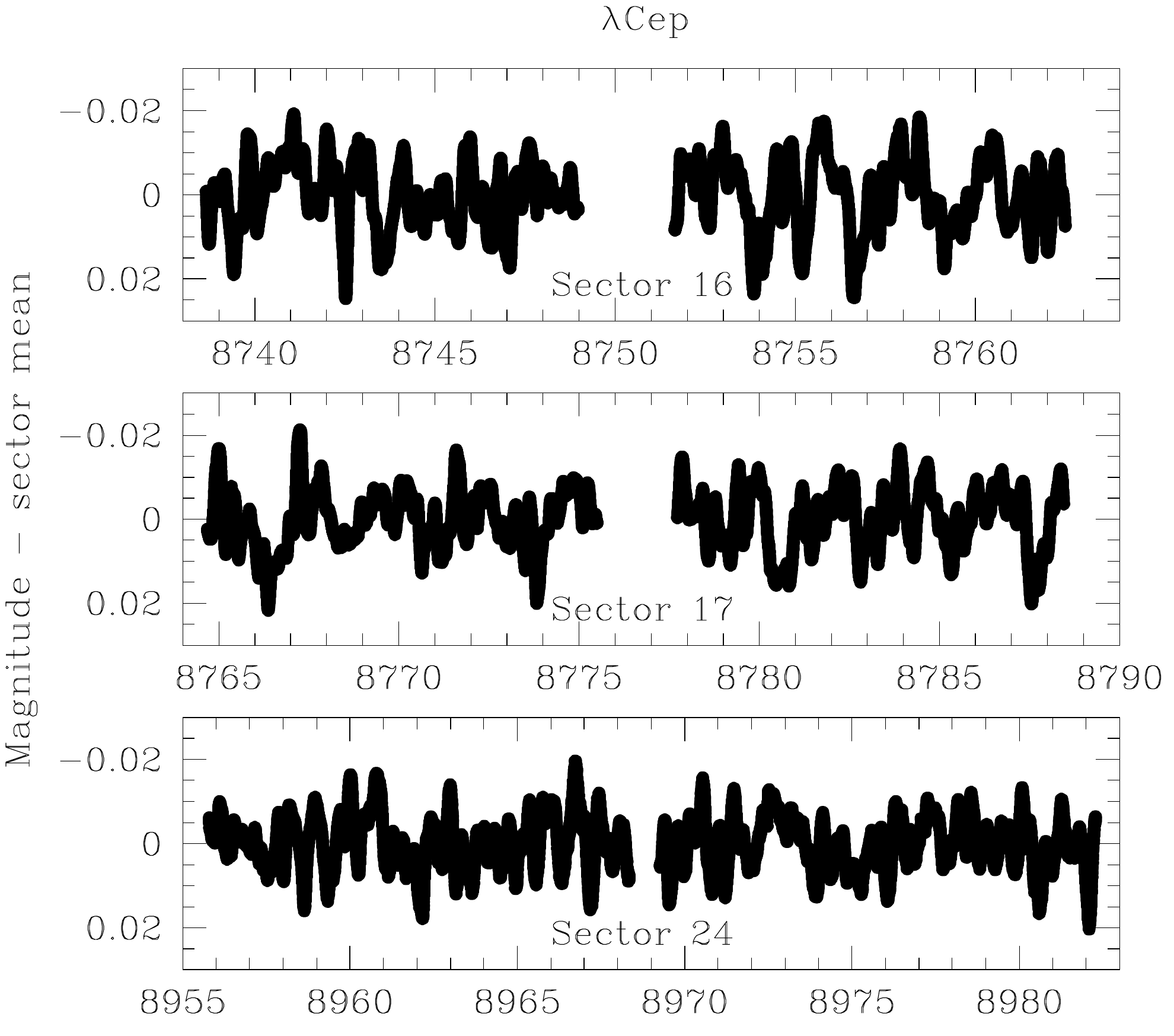}}
  \resizebox{8.5cm}{!}{\includegraphics{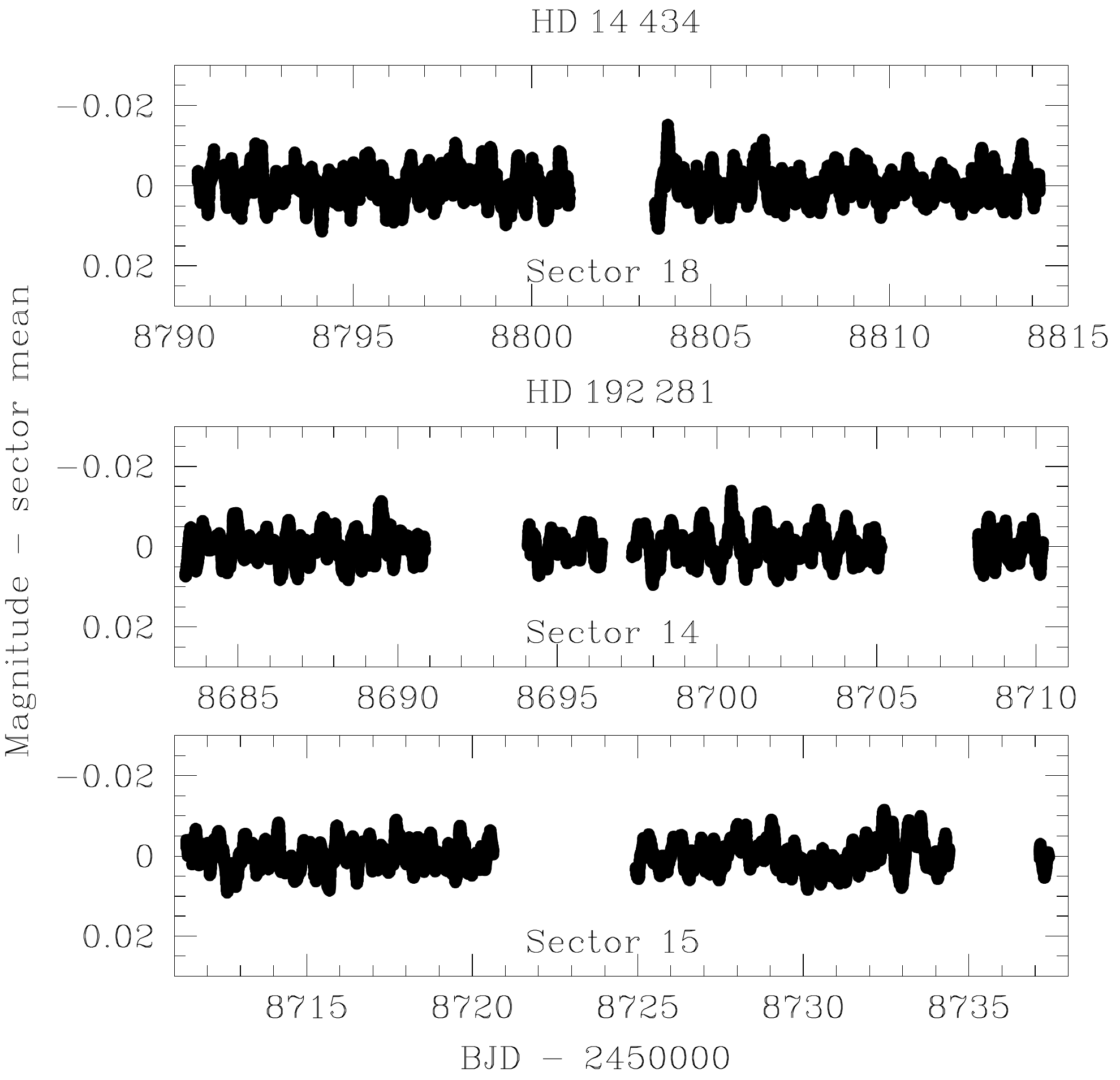}}
\end{center}  
\caption{{\it TESS} 2\,min PDC lightcurves of $\lambda$~Cep, HD~14\,434 and HD~192\,281. All lightcurves are shown after subtraction of the mean magnitude from the data of each sector. \label{lcurve}}
\end{figure}
\begin{figure}
\begin{center}
  \resizebox{8.5cm}{!}{\includegraphics{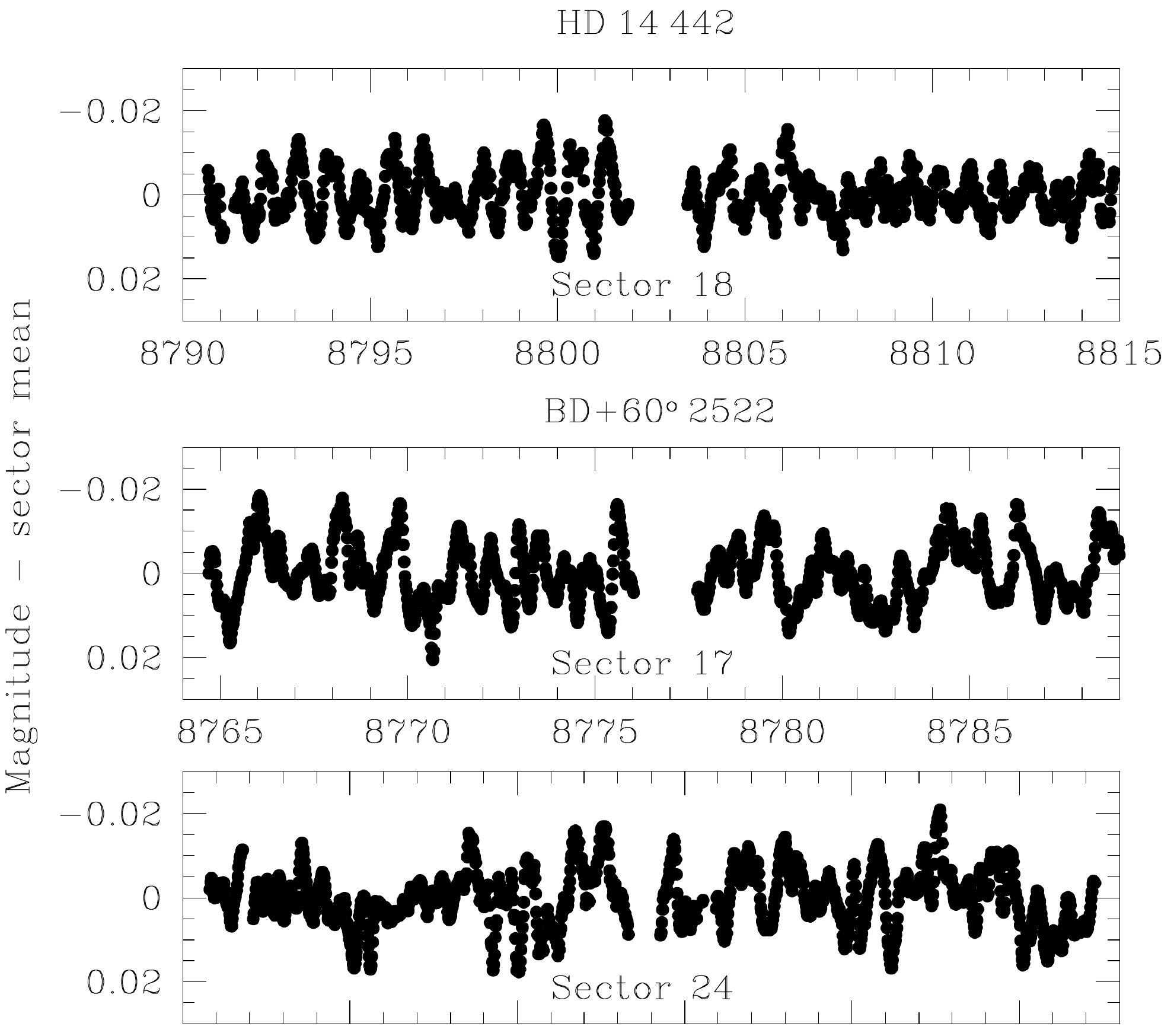}}
  \resizebox{8.5cm}{!}{\includegraphics{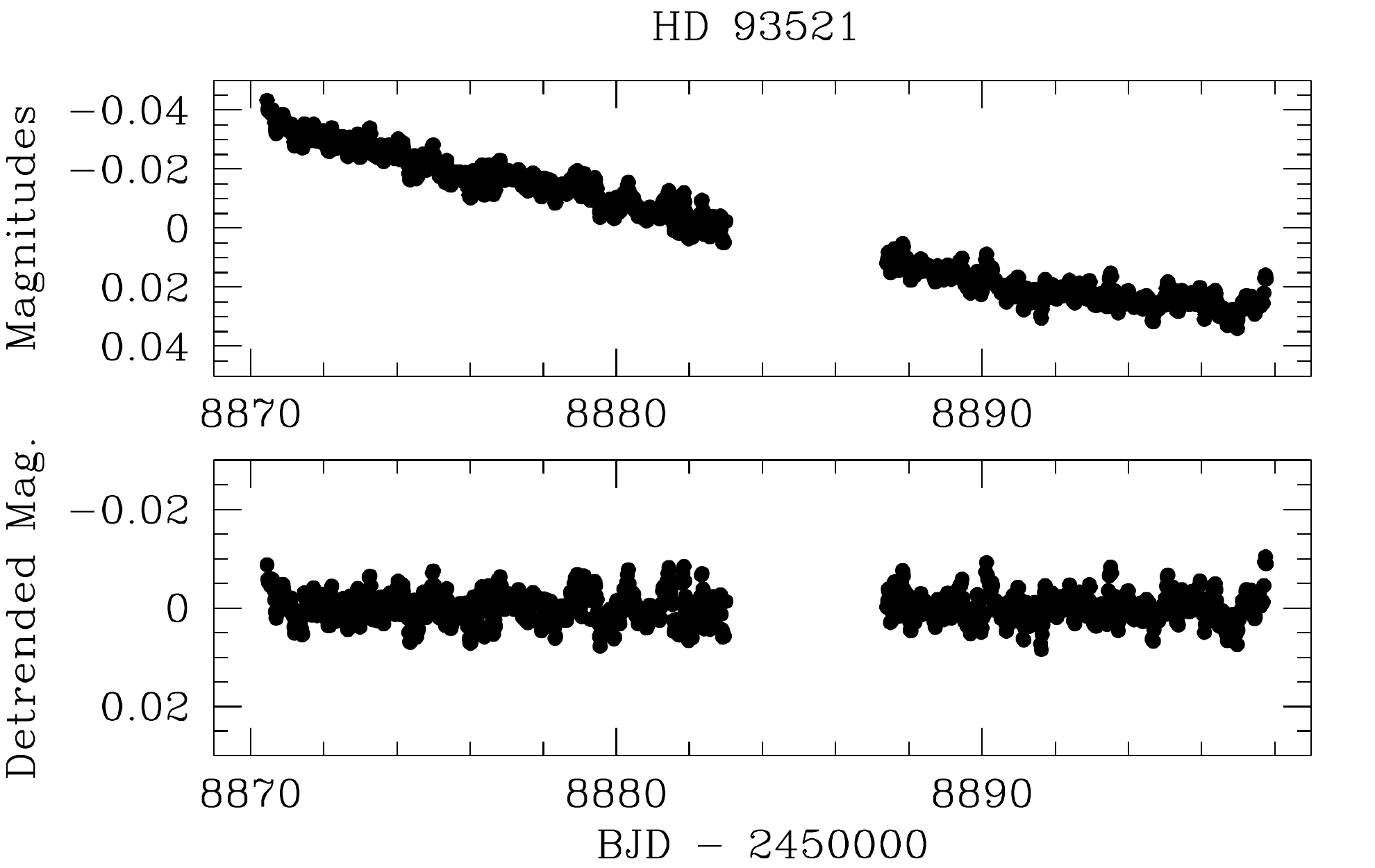}}
\end{center}  
\caption{Top three panels: FFI 30\,min cadence {\it TESS} lightcurves of HD~14\,442 and BD$+60^{\circ}$\,2522 after subtraction of the mean magnitude of each sector. Bottom two panels: FFI 30\,min cadence {\it TESS} lightcurve of HD~93\,521 during sector 21 after discarding the most deviating points around the transition between the orbits and after removing the long-term trends. \label{lcHD93521}}
\end{figure}

For HD~14\,442, BD$+60^{\circ}$\,2522 and HD~93\,521, the full frame images (FFI) with a 30\,min cadence (Nyquist frequency of 24\,d$^{-1}$) were processed with the Python software package Lightkurve\footnote{https://docs.lightkurve.org/}. Aperture photometry was extracted on $50 \times 50$ pixels image cutouts. For the source mask, we adopted a flux threshold of 2, 10, or 15 times the median absolute deviation over the median flux for HD~93\,521, HD~14\,442, and BD$+60^{\circ}$\,2522, respectively. The background level was evaluated from the pixels in the $50 \times 50$ pixels cutout that were below the median flux, and background subtraction was performed by means of a principal component analysis including 5 components. The fluxes were converted to magnitudes and data points with errors exceeding the mean error augmented by three times the dispersion of the errors were removed. The formal accuracies of the FFI photometric data are 0.14\,mmag, 0.10\,mmag and 0.07\,mmag respectively for HD~14\,442, BD$+60^{\circ}$\,2522, and HD~93\,521. 

The {\it TESS} lightcurves of our Onfp/Oef targets are illustrated in Figs.\,\ref{lcurve} and \ref{lcHD93521}. All stars are found to display variability, although with different amplitudes. The strongest variability is observed for $\lambda$~Cep, HD~14\,442 and BD$+60^{\circ}$\,2522, with variations of 0.04\,mag peak to peak whereas the range is $\sim$0.02\,mag for the other two stars. It is interesting to note that the stars with the strongest variability are those displaying the strongest He\,{\sc ii} $\lambda$~4686 emission lines \citep[see][and Fig.\,\ref{meanspec} below]{Rau03,DeB04} and having the highest luminosities (see Sect.\,\ref{revision}).

Special attention had to be paid to HD~93\,521 (Fig.\,\ref{lcHD93521}). Indeed, the source was located very close to the edge of the field of view of camera 1, rendering the data reduction more complex. Because of the proximity of the edge of the field of view, some part of the flux of the source was lost. This led to apparent long-term variations (Fig.\,\ref{lcHD93521}). Furthermore, data points around the mid-sector gap (BJD-2\,450\,000 between 8883 and 8887.3) displayed larger and faster variations hence were discarded. To correct for the trend, we used two different approaches. We either fitted the data from each orbit with a second order polynomial, or applied a 2-day sliding average window to determine the trend. These presumably instrumental variations were then subtracted from the observations (see bottom panel of Fig.\,\ref{lcHD93521}). Both detrending methods yielded very similar results. In the following, only the data detrended with the 2-day sliding window method are presented.

Given the size of the {\it TESS} pixels and the fact that the flux of a given source is extracted over apertures of several pixels, crowding can become an issue. We have thus checked the {\it GAIA}-DR2 catalog \citep{Brown} for the magnitudes of neighbouring sources within a radius of 1\arcmin\ around our targets. In the worst case (HD~14\,434), the brightest contaminating source is 4.4\,mag fainter than our target. In all other cases, the brightest neighbours are between 5.3 and 11\,mag fainter than the stars of interest. We further note that the {\it GAIA}-DR2 astrometric solutions of all our targets have a Renormalised Unit Weight Error (RUWE) between 0.85 and 1.11. Hence, there is no indication of non-single sources or otherwise problematic astrometric solutions.
Therefore, we conclude that contamination of the {\it TESS} photometry due to source confusion should be negligible for all our targets.

\begin{figure*}
  \begin{center}
    \begin{minipage}{8.5cm}
      \resizebox{8.5cm}{!}{\includegraphics{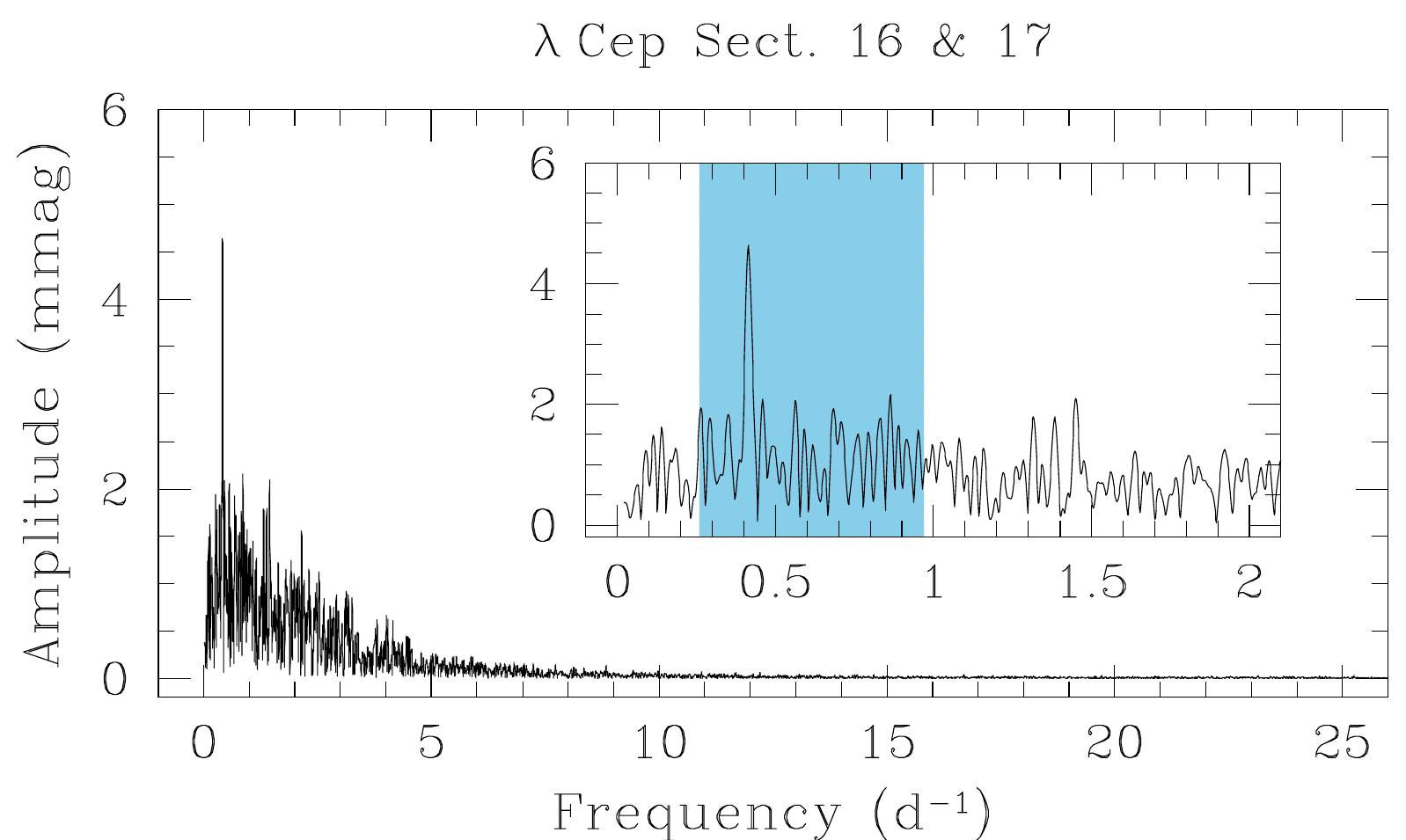}}
      \resizebox{8.5cm}{!}{\includegraphics{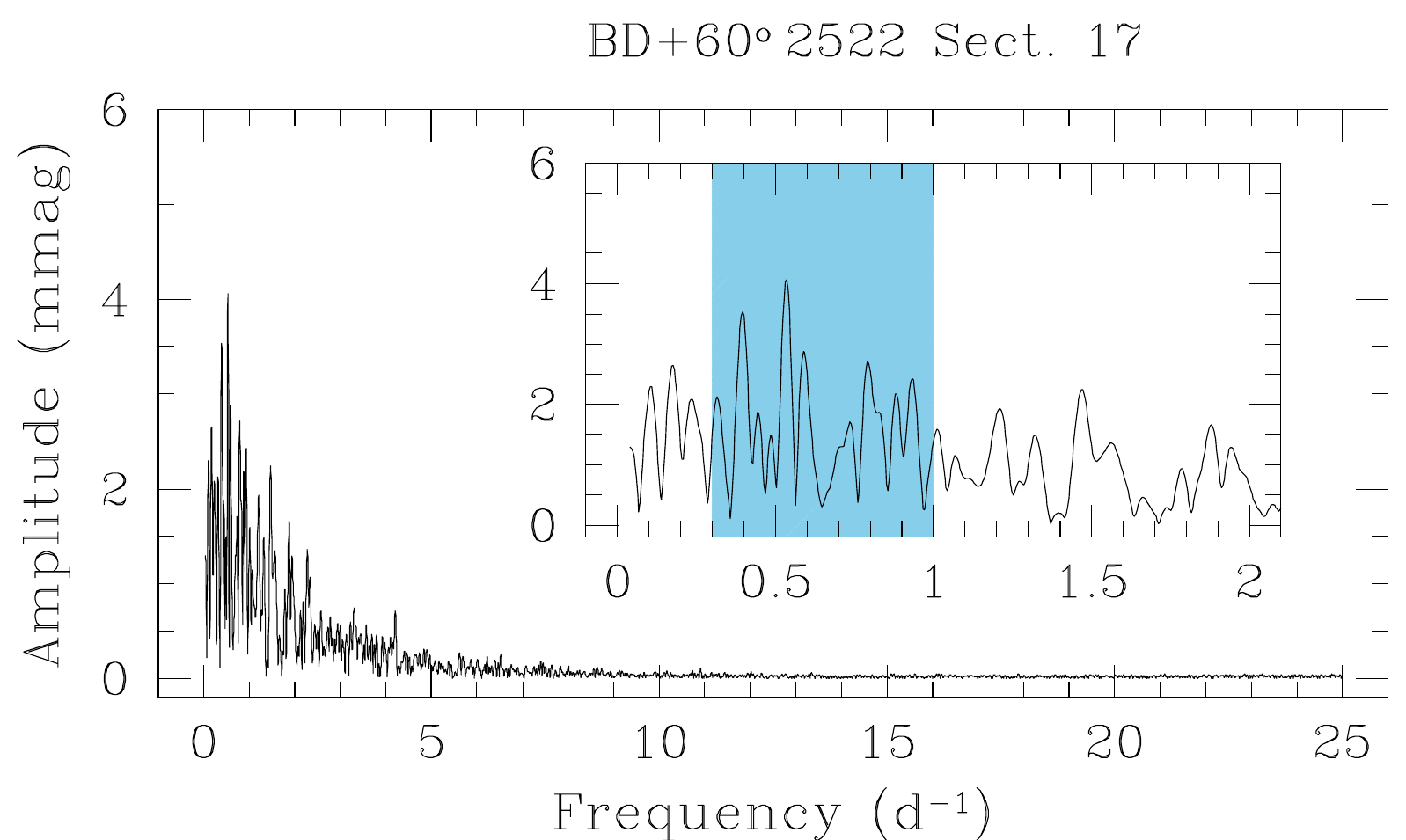}}
      \resizebox{8.5cm}{!}{\includegraphics{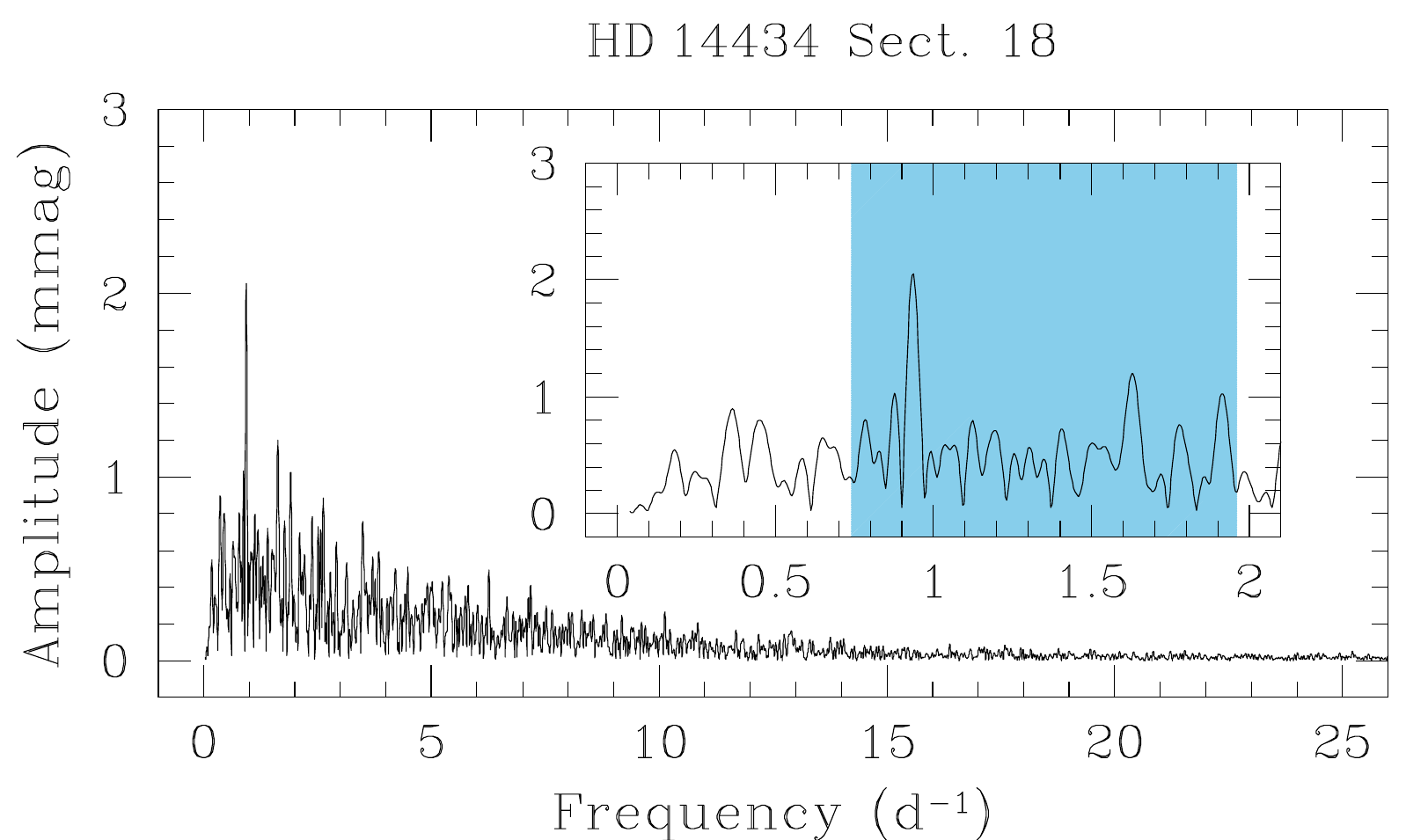}}
      \resizebox{8.5cm}{!}{\includegraphics{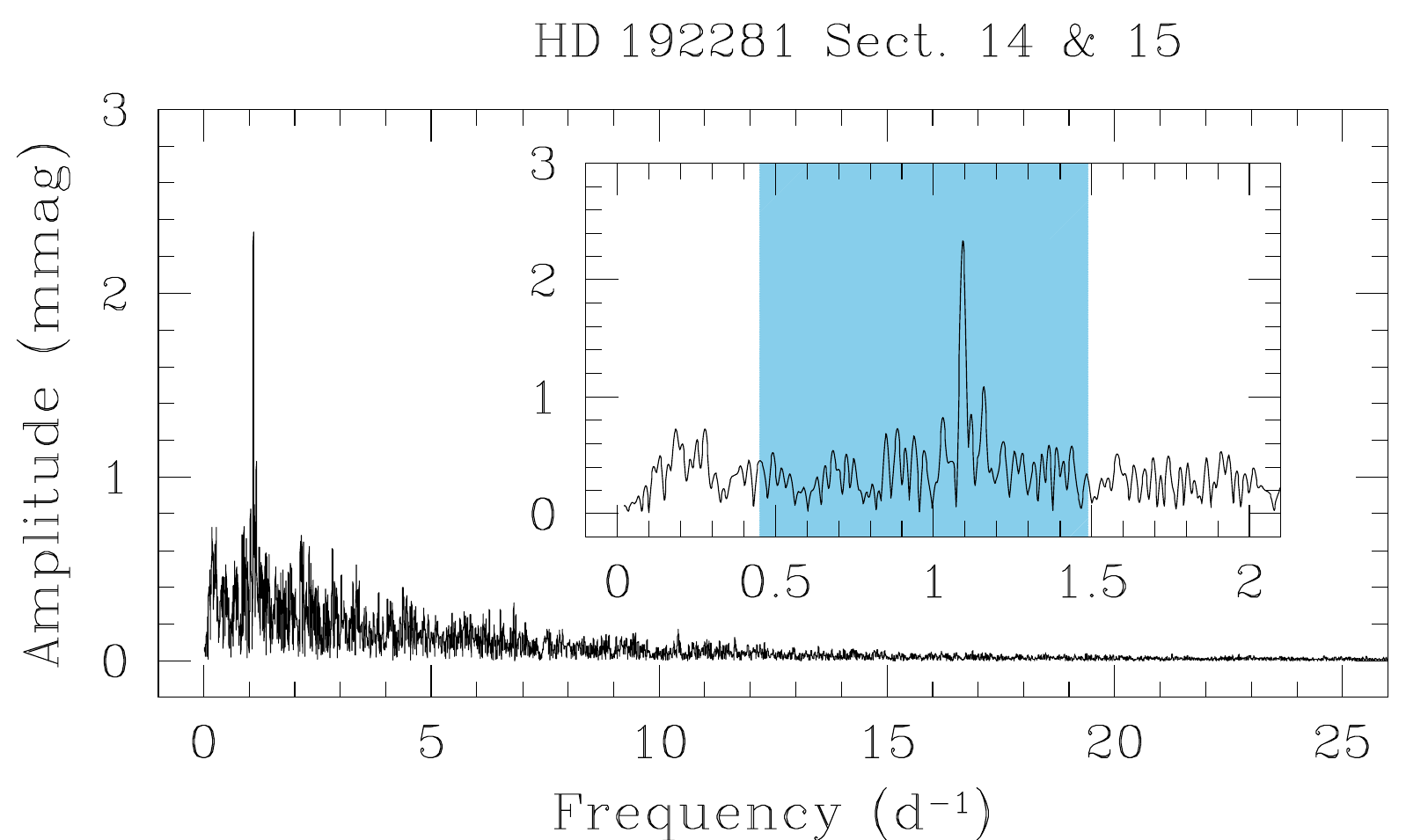}}
    \end{minipage}
    \hfill
    \begin{minipage}{8.5cm}
      \resizebox{8.5cm}{!}{\includegraphics{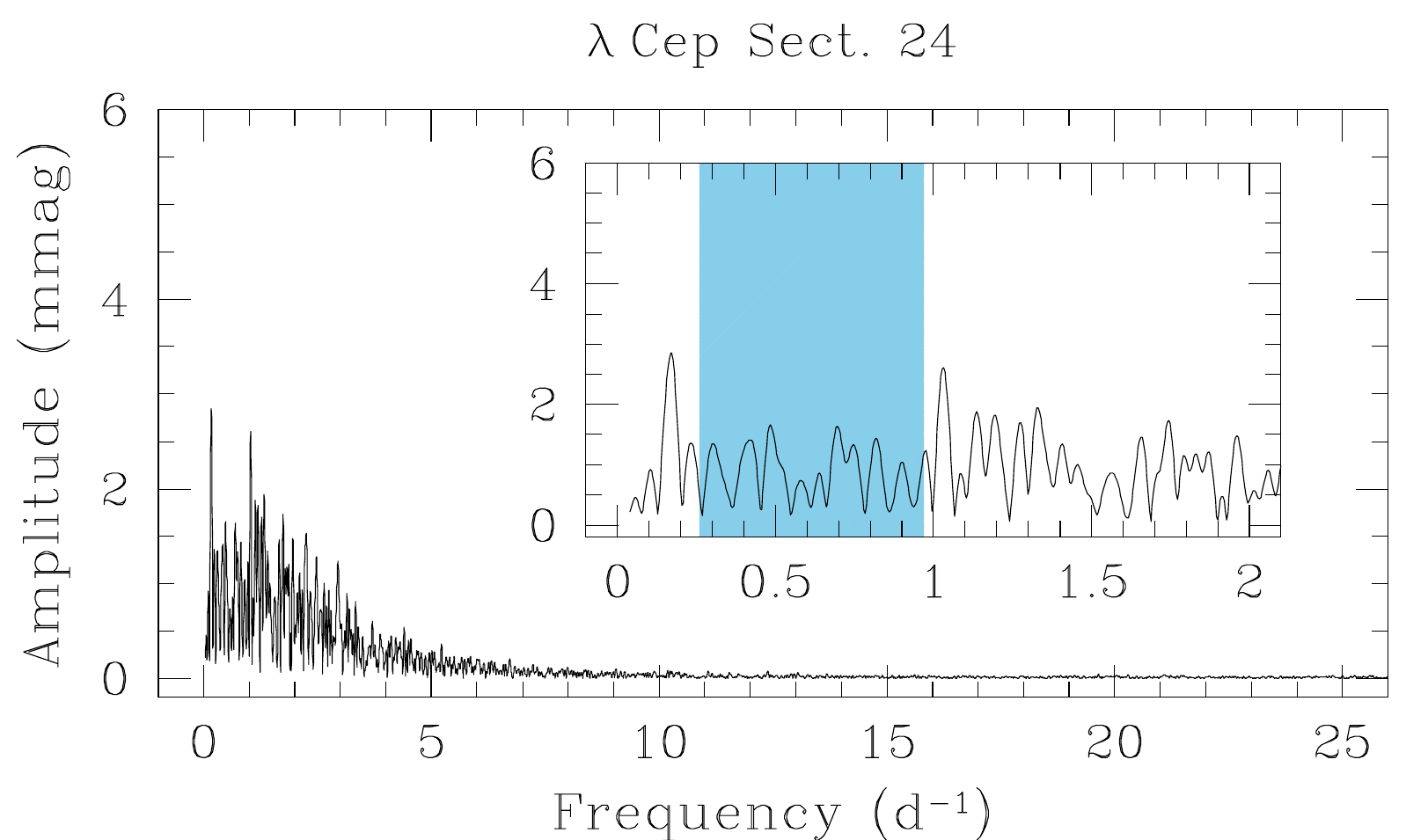}}
      \resizebox{8.5cm}{!}{\includegraphics{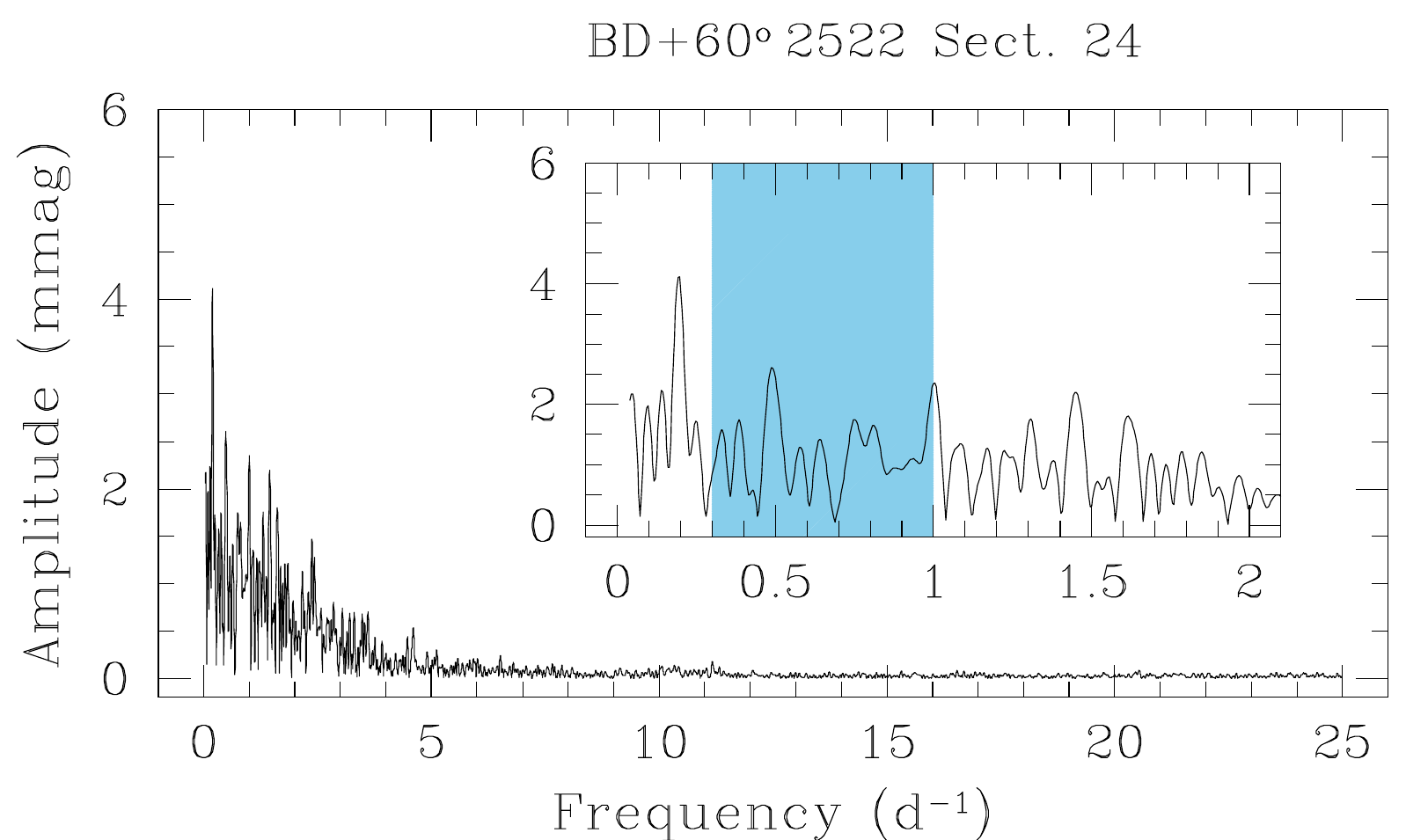}}
      \resizebox{8.5cm}{!}{\includegraphics{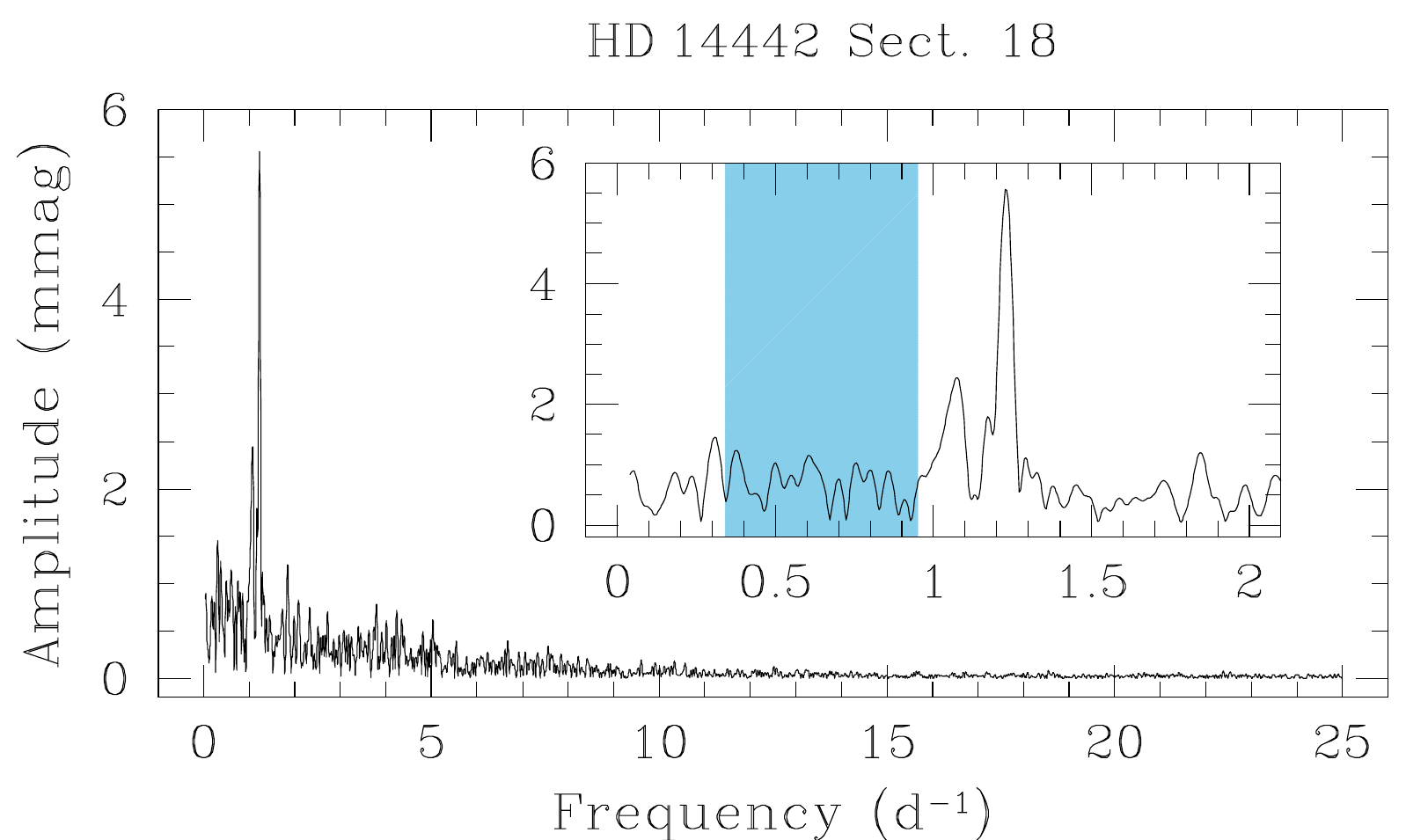}}
      \resizebox{8.5cm}{!}{\includegraphics{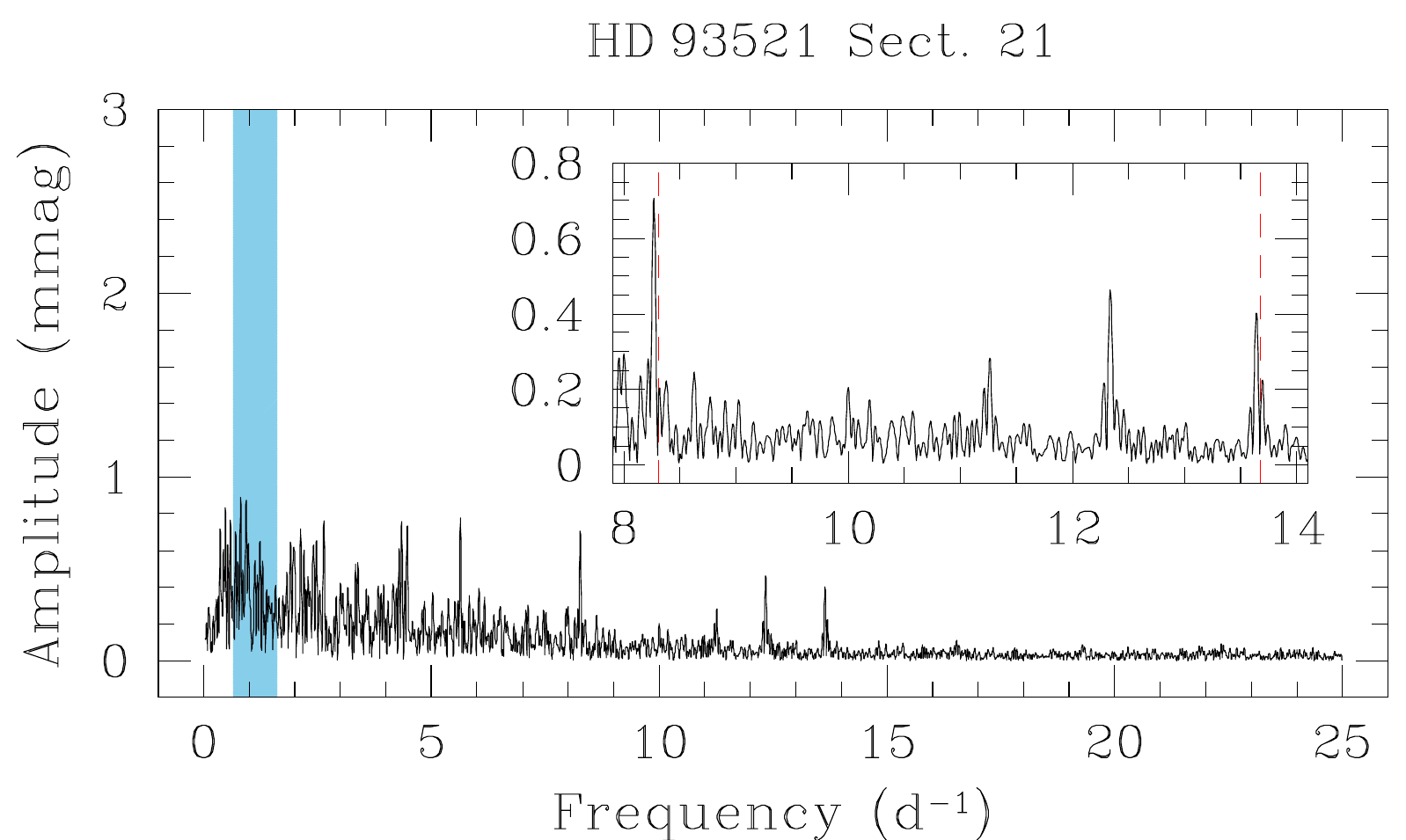}}
    \end{minipage}
  \end{center}
  \caption{Fourier periodograms of the {\it TESS} photometry of the Onfp/Oef stars and HD~93\,521. The inserts illustrate a zoom on the low-frequency range, except for HD~93\,521 where the zoom is on the region around the frequencies associated with non-radial pulsations (red dashed lines) detected in optical spectroscopy. The blue shaded areas indicate the range of frequencies compatible with $\nu_{\rm rot}$ (see Table\,\ref{keyprop2}).\label{periodogram}}
\end{figure*}

\subsection{Spectroscopy}
A series of 76 blue spectra of $\lambda$~Cep were collected with the Aur\'elie spectrograph \citep{Gillet} at the 1.52\,m telescope of the Observatoire de Haute Provence (OHP; France). The data were taken during six observing nights between 2 and 7 June 2015, resulting in a natural width of the peaks in the periodogram of $\Delta \nu_{\rm nat,OHP} = 0.19$\,d$^{-1}$. Aur\'elie was equipped with a $2048 \times 1024$\,CCD with a pixel size of 13.5\,$\mu$m squared. A 1200\,l\,mm$^{-1}$ grating, blazed at 5000\,\AA, provided a resolving power of 14\,200 over the wavelength domain from 4580\,\AA\ to 4770\,\AA\ as measured on Th-Ar lamp exposures. Typical integration times were 20\,min, and the mean time interval between the middle of two consecutive exposures of a given night was $22$\,min = $0.0156$\,d, implying a Nyquist frequency of $32.1$\,d$^{-1}$. The mean S/N ratio, evaluated in the line-free region between 4735 and 4745\,\AA, was 520. Hollow cathode Th-Ar lamp exposures were taken every hour and a half during the nights. The data were reduced using version 17FEBpl\,1.2 of the {\sc midas} software developed at ESO. For each observation, the Bowen-Walraven image slicer of the Aur\'elie spectrograph produces five sub-images of the spectrum in the focal plane. Each of these sub-images was independently extracted and flat-field corrected. This procedure allowed removing cosmic rays before combining the sub-images into a single spectrum. The typical rms error of the wavelength solution was 0.0027\,\AA. The normalization was perfomed using a set of line-free continuum windows. The normalized spectra are made available on the journal website.

\section{Analysis}\label{analysis}
\subsection{Overall periodogram} \label{TESSanalyse}
The time series of all targets were analysed with the modified Fourier periodogram algorithm of \citet{HMM} and \citet{Gosset}, which explicitly accounts for uneven sampling. For those stars where more than one sector of observations is available, we first analysed the data from each sector separately then data from the same star and from consecutive sectors were combined and the Fourier analysis was repeated. 
 
Unlike the case of HD~93\,521, the periodograms of the {\it TESS} photometry of Onfp/Oef stars reveal no significant signal at frequencies above $\sim 6$\,d$^{-1}$ (Fig.\,\ref{periodogram}). The only possible exception are the data from BD$+60^{\circ}$\,2522 during sector 24, where a pair of low-level peaks (or groups of peaks) is seen at 10.3 and 11.2\,d$^{-1}$ with amplitudes of 0.20\,mmag and 0.21\,mmag, though these peaks are absent from the periodogram of the same star observed during sector 17. Overall, the periodograms clearly exhibit the signature of red noise. This terminology is used for apparently stochastic signals whose power increases towards lower frequencies (see Sect.\,\ref{rednoise}). In general terms, the periodograms of the Onfp/Oef stars can be put into two broad categories: those displaying red noise without a clearly dominant peak (i.e.\ no peak has an amplitude exceeding three times the level of the red and white noise) and those displaying an outstanding peak on top of the red noise continuum (i.e.\ the highest peak has an amplitude of more than four times the level of the red and white noise). In the first category, we find BD$+60^{\circ}$\,2522 (sect.\ 17). The second category gathers the periodograms of $\lambda$~Cep (sects.\ 16 \& 17), HD~14\,434, HD\,14\,442, and HD\,192\,281, as well as that of $\zeta$~Pup \citep[][see also Sect.\,\ref{compazetaPup}]{Tahina}. The cases of $\lambda$~Cep (sect.\ 24) and BD$+60^{\circ}$\,2522 (sect.\ 24) are somewhat intermediate between the two categories.

\begin{figure}
\begin{center}
  \resizebox{9cm}{!}{\includegraphics{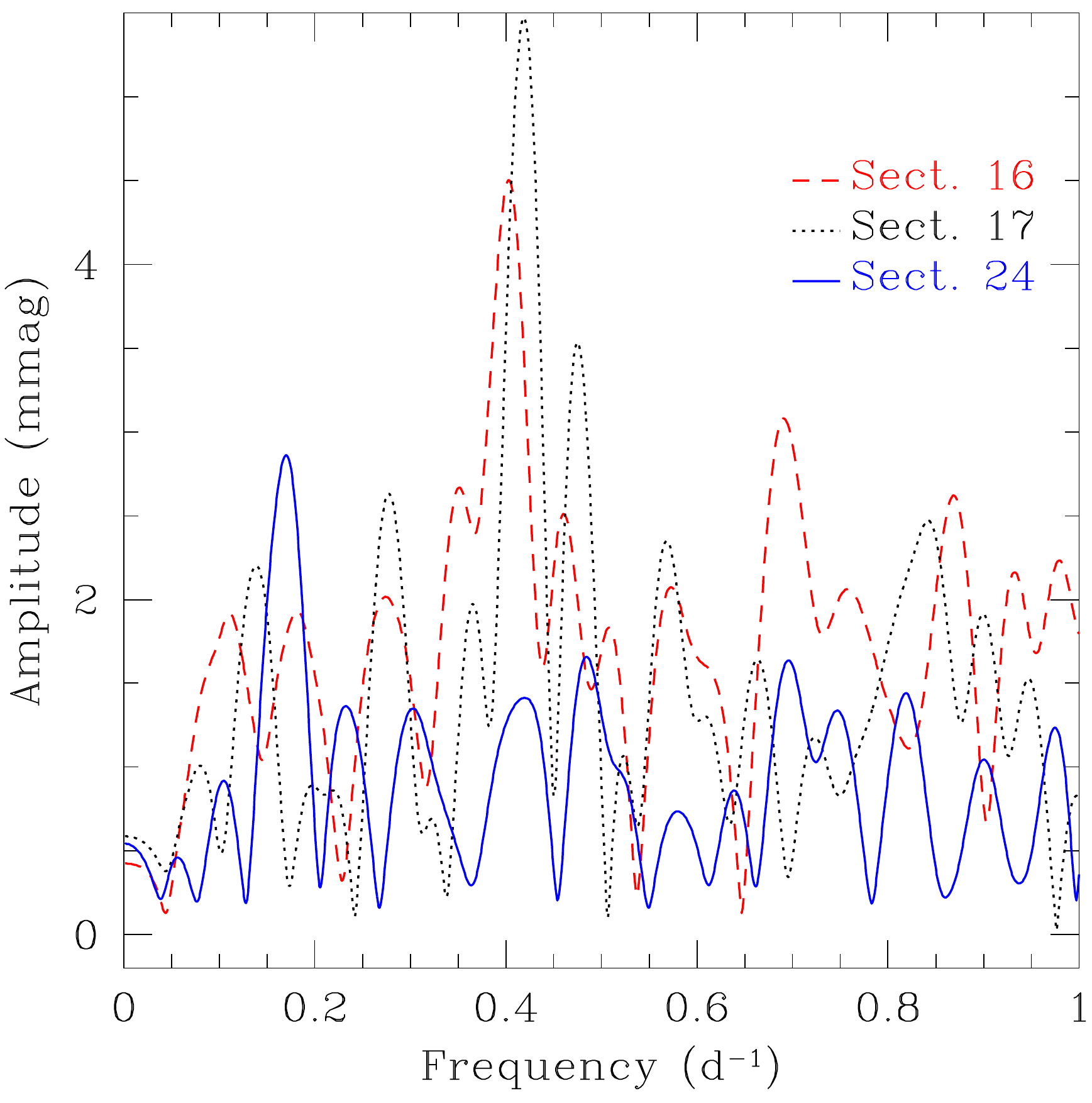}}
\end{center}  
\caption{Comparison between the periodograms built from the $\lambda$~Cep data of sector 16 (dashed red line), sector 17 (dotted black line) and sector 24 (blue solid line).\label{zoomperiodogram}}
\end{figure}

An important result is that the same star can switch category, as clearly illustrated by the case of $\lambda$~Cep which displayed a peak at $\nu_{{\rm TESS,}~\lambda~{\rm Cep}} = 0.414 \pm 0.002$\,d$^{-1}$ with an amplitude of $4.64$\,mmag in the combined dataset from sectors 16 and 17, whilst no such peak was present six months later. Figure\,\ref{zoomperiodogram} provides a zoom on the low-frequency part of the periodograms of the $\lambda$~Cep data from the individual sectors. In the power spectrum of the data from sector 16, the strongest peak is found at a frequency of $0.403 \pm 0.004$\,d$^{-1}$ with an amplitude of $4.50$\,mmag. For the data from sector 17, the strongest peak occurs at $0.419 \pm 0.004$\,d$^{-1}$ with an amplitude of $5.47$\,mmag. Whilst the profiles of both peaks overlap, their centroids clearly differ (Fig.\,\ref{zoomperiodogram}). Furthermore, in the data from sector 24, the peak at $0.414$\,d$^{-1}$ has totally disappeared (see Fig.\,\ref{zoomperiodogram}). Instead, the strongest signal is now found at a frequency of $0.170 \pm 0.004$\,d$^{-1}$ but, with an amplitude of 2.9\,mmag, it barely stands out against the red noise. Indeed, a second peak of nearly identical strength (amplitude 2.6\,mmag) is found at $1.031 \pm 0.004$\,d$^{-1}$ (Fig.\,\ref{periodogram}).

To further investigate how the frequency content evolves with time, we computed time-frequency diagrams by performing a Fourier analysis of the photometric time series cut into sliding windows of a duration of ten days and with a step of 1\,day. The resulting time-frequency diagrams of $\lambda$~Cep are presented in Fig.\,\ref{spevol}. Similar plots are shown for the other stars in Fig.\,\ref{FigAppen1} of Appendix\,\ref{Appendix1}.
\begin{figure*}
  \begin{minipage}{8.5cm}
    \begin{center}
      \resizebox{8.5cm}{!}{\includegraphics{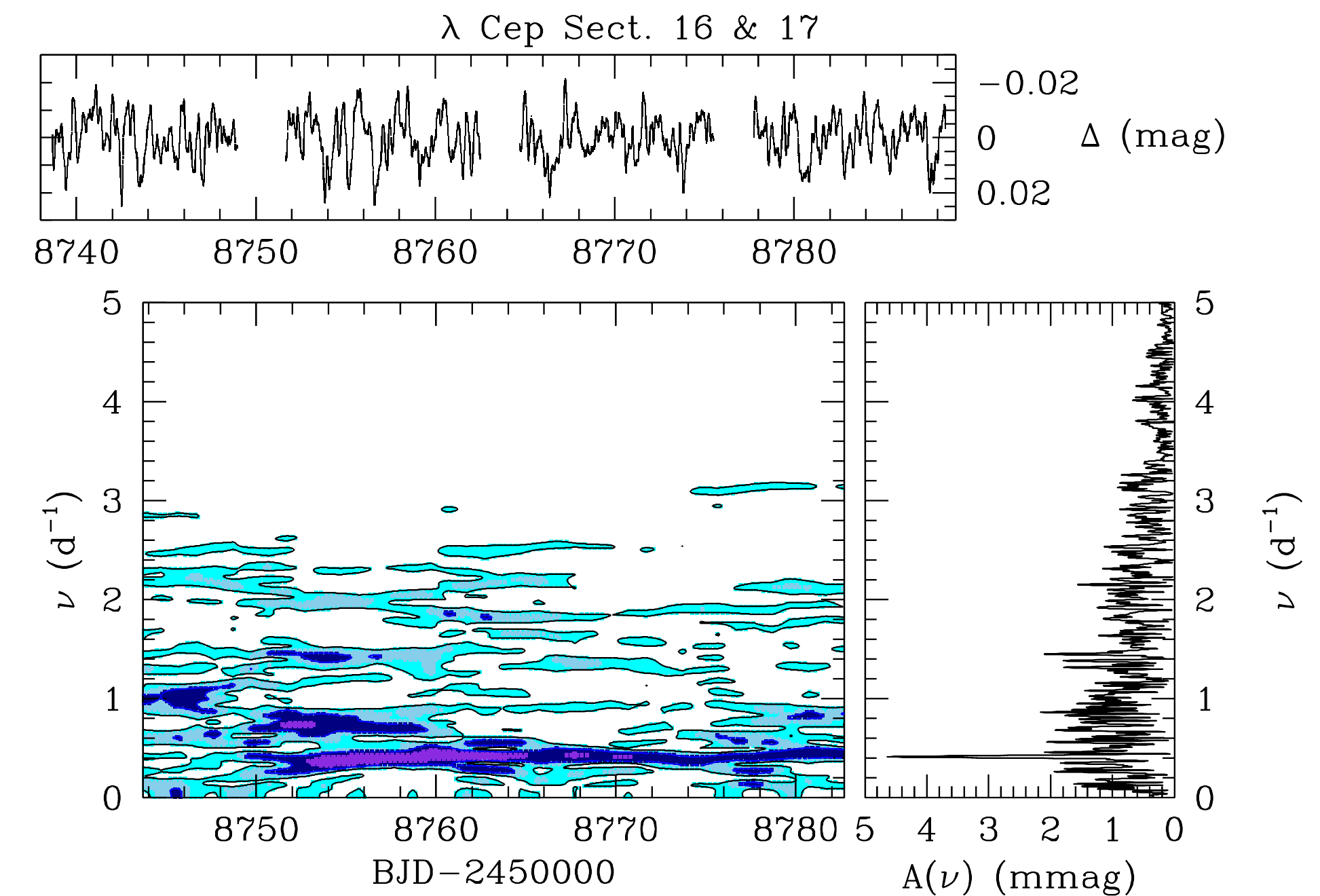}}
    \end{center}
  \end{minipage}
  \hfill
  \begin{minipage}{8.5cm}
    \begin{center}
      \resizebox{8.5cm}{!}{\includegraphics{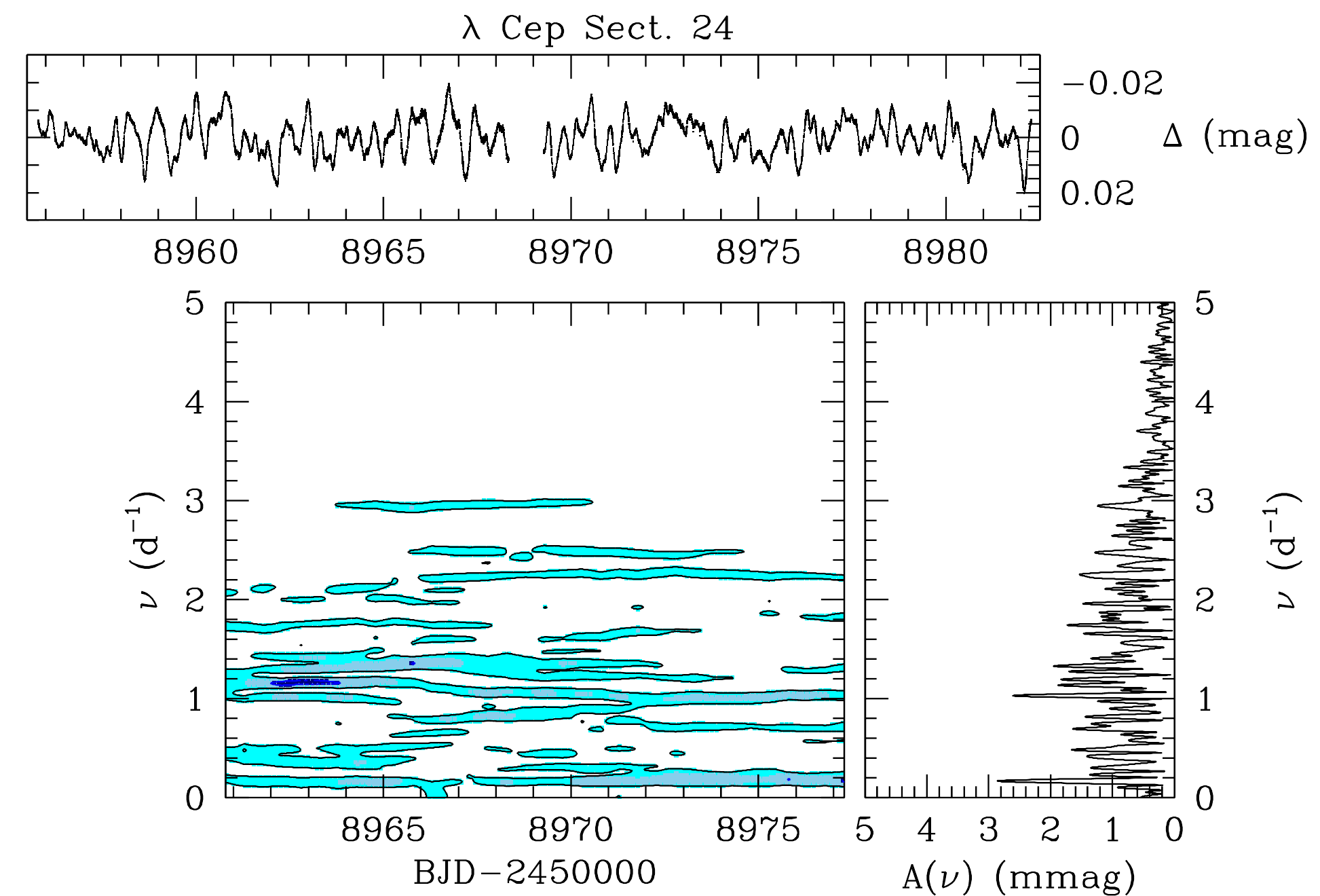}}
    \end{center}
  \end{minipage}  
  \caption{{\it Left}: time-frequency diagrams of the sectors 16 \& 17 {\it TESS} photometric time series of $\lambda$~Cep. The top panel shows the observed lightcurve. The bottom left panel provides the evolution of the Fourier periodogram with the date corresponding to the middle of the 10-day sliding window. Violet, dark blue, light blue, and cyan colours stand respectively for amplitudes $\geq 6$\,mmag, $\geq 4$\,mmag, $\geq 3$\,mmag and $\geq 2$\,mmag. The right vertical panel illustrates the Fourier periodogram evaluated over the full duration of sectors 16 and 17. {\it Right}: same for the {\it TESS} observations from sector 24. \label{spevol}}
\end{figure*}

The visibility of the $0.414$\,d$^{-1}$ peak clearly changes with time. It was nearly absent until BJD\,2\,458\,752, reached its maximum strength between BJD\,2\,458\,754 and 2\,458\,764 and remained visible, though with a somewhat reduced amplitude, until the end of sector 17. Moreover, there is a slight variation of the peak frequency with time: the frequency first increased and then decreased, before it increased again towards the end of the campaign. This behaviour is responsible for the shift between the peaks of sectors 16 and 17 seen in Fig.\,\ref{zoomperiodogram}. These properties are suggestive of a transient feature rather than of a persistent signal.

Quite similar descriptions apply to the Fourier analyses of the data of the other Onfp/Oef stars. BD$+60^{\circ}$\,2522 appears as the most extreme example (Fig.\,\ref{FigAppen1}): not only did its periodogram change tremendously between sectors 17 and 24, but during a sector, all peaks, with the exception of the 0.196\,d$^{-1}$ frequency that persisted during the whole sector 24, have lifetimes under ten days. At the other end, the most stable signal is the strong modulation seen in HD~14\,442 (Fig.\,\ref{FigAppen1}), though even in this case we observe slight changes of the amplitude over the duration of sector 18.

The Fourier periodogram of the non-radial pulsator HD~93\,521 (Fig.\,\ref{periodogram}) is dominated by some excess power at frequencies below 5\,d$^{-1}$ which is reminiscent of the red noise seen in the Onfp/Oef stars and in many other OB stars \citep[e.g.][]{Bow20}. In view of the flux losses due to the proximity of the source to the edge of the field of view, one cannot rule out the possibility that some part of the low-frequency content of the power spectrum might stem from instrumental effects. However, not everything is of instrumental origin. Indeed, low frequency excess power was also seen in the spectroscopic data of HD~93\,521 by \citet{Rau08}. In addition, we note the presence of discrete peaks at frequencies of 5.63, 8.27, 12.34 and 13.64\,d$^{-1}$. Whilst the first, and third of these frequencies have no known spectroscopic counterparts, the second and the fourth correspond to the frequencies $\nu_2$ and $\nu_1$ that rule the line profile variability in the optical spectrum and were attributed to NRP modes with $\ell \simeq 4 \pm 1$ and $\ell \simeq 8 \pm 1$, respectively \citep{Rau08}. The amplitudes of these modes are $0.71 \pm 0.08$\,mmag and $0.40 \pm 0.02$\,mmag. The time-frequency diagram (Fig.\,\ref{FigAppen1}) reveals that the frequencies of the four signals identified above are stable over the duration of the {\it TESS} sector. However, as becomes clear from Fig.\,\ref{FigAppen1}, even in the case of persistent pulsation modes, one can observe modulations of their amplitudes. We shall come back to this point in Sect.\,\ref{discussion}.

\begin{table*}
  \caption{Strongest peaks and red noise properties of the periodograms of Onfp/Oef stars \label{tab:periodogram}}
  \begin{tabular}{c c c c c c c c c}
    \hline\hline
    Star & Sector & \multicolumn{2}{c}{Strongest peak(s)} & \multicolumn{4}{c}{Red noise} \\
    \cline{3-4}\cline{6-9}
    &    & $\nu$ & $A_{\nu}$ &&$A_0$ & $\tau$ & $\gamma$ & $C_{\rm white\,noise}$ \\
    &    & (d$^{-1}$) & (mmag) && (mmag) & (d) &          & (mmag) \\
    \hline
    $\lambda$~Cep & 16 \& 17 & 0.414 & 4.64 && $0.97 \pm 0.01$ & $0.060 \pm 0.001$ & $2.76 \pm 0.04$ & $0.009 \pm 0.001$ \\
                  & 24 & 0.170,(1.031) & 2.86,2.61 && $0.93 \pm 0.01$ & $0.050 \pm 0.001$ & $3.60 \pm 0.08$ & $0.013 \pm 0.001$ \\
    \hline
    HD~14\,434 & 18 & 0.936 & 2.05 && $0.42 \pm 0.01$ & $0.030 \pm 0.001$ & $2.26 \pm 0.05$ & $0.010 \pm 0.001$\\
    \hline
    HD~14\,442 & 18 & 1.230,1.075 & 5.58,2.45 && $0.72 \pm 0.04$ & $0.054 \pm 0.004$ & $1.52 \pm 0.17$ & $\leq 0.016$\\
    \hline
    HD~192\,281 & 14 \& 15 & 1.093 & 2.34 && $0.33 \pm 0.01$ & $0.037 \pm 0.001$ & $2.10 \pm 0.03$ & $0.005 \pm 0.001$\\
    \hline
    BD$+60^{\circ}$\,2522 & 17 & (0.535) & 4.07 && $1.67 \pm 0.07$ & $0.097 \pm 0.006$ & $2.12 \pm 0.19$ & $0.011 \pm 0.013$\\
    & 24 & 0.194 & 4.13 && $1.20 \pm 0.04$ & $0.070 \pm 0.003$ & $2.81 \pm 0.24$ & $0.027 \pm 0.009$ \\
    \hline
  \end{tabular}
  
  {\scriptsize The frequencies and amplitudes in the third and fourth columns are listed by decreasing amplitude. Frequencies between brackets correspond to peaks that stand out only marginally above the level of the red noise, i.e.\ with an amplitude ratio of less than 3.}
\end{table*}

\subsection{Red noise}\label{rednoise}
All periodograms displayed the presence of red noise. Several tests were performed to check whether it might be of instrumental or processing origin. First, the Fourier periodograms of the simple aperture photometry were compared to those of the PDC photometry \citep[corrected for the known instrumental trends,][]{Jen16}. Apart for some slight differences at frequencies below 0.1\,d$^{-1}$, the periodograms from both sets of photometric data are in excellent agreement. Second, data taken within 0.3\,d of the beginning or end of the observations (including around the mid-sector break and between the sectors) were discarded to get rid of putative instrumental sensitivity drops. The resulting periodograms were essentially identical to those obtained with the full set of PDC data of $\lambda$~Cep. Finally, some stars have been observed in the same sector and with the same camera but yield different low-frequency features despite the same instrument being used for all. Thus, no strong indication for the presence of instrumental red noise emerges from these tests.   

Red noise of stellar origin has been found in the photometric times series of a number of early-type stars observed with different facilities. Indeed, red noise was reported in {\it CoRoT} photometry of the O4\,V((f$^+$)) star HD~46\,223, the O5.5\,V((f)) star HD~46\,150, and the O8\,V star HD~46\,966 \citep{Blomme}, {\it Kepler} photometry of the supergiants HD~188\,209 \citep[O9.5\,Iab,][]{Aerts} and HD~91\,316 \citep[B1\,Iab,][]{Aerts18}, as well as in {\it BRITE} photometry of the O4\,Ief supergiant $\zeta$~Pup \citep{Tahina} and of the O7.5\,I(f) + ON9.7\,I binary HD~149\,404 \citep{Rau19}. Red noise was also found in {\it Kepler K2} and {\it TESS} photometry of a large set of Galactic and LMC OB stars \citep{Bow19,Bow20}, and in {\it TESS} photometry of Be and Wolf-Rayet stars \citep{Naze20,Naze21}. Moreover, a red noise component has been previously reported in the variability of the He\,{\sc i} $\lambda$~4471 absorption line of $\lambda$~Cep \citep{Uuh14}, and in the spectroscopic variability of several late OB supergiants \citep[e.g.\ HD~2\,905, B1\,Ia,][]{SD18}. Among the scenarios that have been proposed to explain red noise variability of stellar origin, there are internal gravity waves \citep[see][and references therein]{Tahina,Bow19,Bow20}. Whether these $g$ modes are generated in a subsurface convection zone \citep{Cant,Gra15} or directly in the convective core of the  massive star \citep{Rog13,AR15,Bow19,Bow20} remains currently controversial. \citet{Bow19,Bow20} argue that the global properties of the red noise components, the correlation between the level of macroturbulence and the amplitude of the red noise, as well as the insensitivity of the red noise to metallicity tend to favour the scenario where internal gravity waves are generated directly in the convective core. However, \citet{Lecoanet} found that the wave transfer function, which links the amplitude of $g$ modes generated in the convective core to the surface brightness variation, displays regularly spaced peaks near frequencies of 1\,d$^{-1}$, unlike the observed red noise spectra. These authors thus argue that the observed low-frequency photometric variability in massive stars is more likely to arise from sub-surface convective layers.

\begin{figure}
\begin{center}
\resizebox{8cm}{!}{\includegraphics{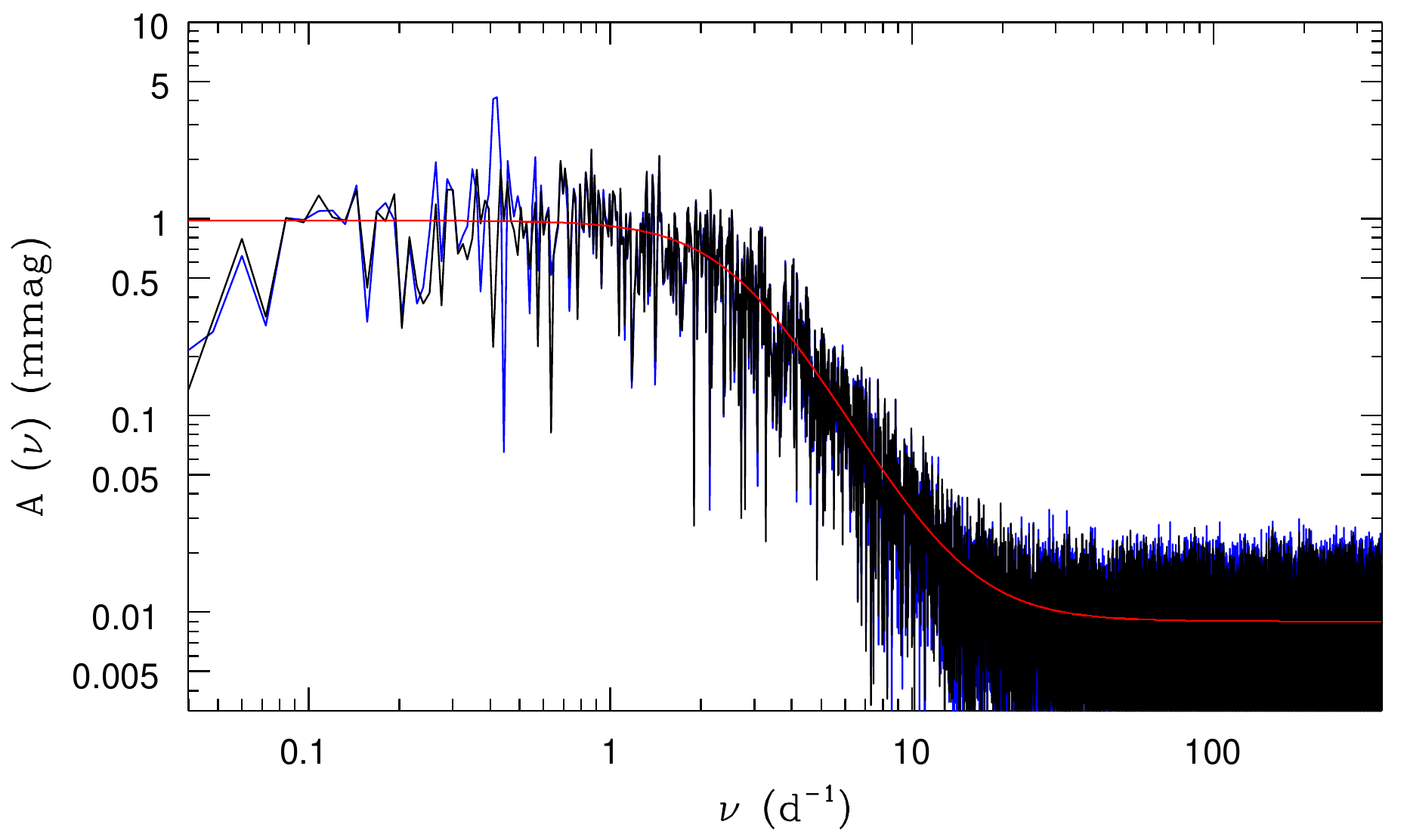}}
\end{center}  
\caption{Log-log plot of the amplitude spectrum of the sector 16 \& 17 {\it TESS} PDC photometry of $\lambda$~Cep. The blue curve illustrates the original power spectrum, whereas the black one corresponds to the power spectrum after prewhitening of the signal at $0.414$\,d$^{-1}$. The red curve corresponds to the best-fit red-noise relation with $A_0 = 0.97$\,mmag, $\gamma = 2.76$, $\tau = 0.060$\,day and $C_{\rm white\,noise} = 0.009$\,mmag, adjusted to the prewhitened power spectrum.}
\label{rnfit}
\end{figure}

Following the same approach as \citet{Blomme} and \citet{Uuh14}, we adopt the formalism of \citet{Stanishev} to fit the red-noise part of the periodograms of our targets: 
\begin{equation}
  A(\nu) = \frac{A_0}{1 + (2\,\pi\,\tau\,\nu)^{\gamma}} + C_{\rm white\,noise}
  \label{eq1}
\end{equation}
Here $A(\nu)$ is the amplitude (in mmag) at frequency $\nu$ in the periodogram inferred from the observations. The scaling factor $A_0$, the slope $\gamma$, the mean lifetime $\tau$ (in days), as well as the level of white noise $C_{\rm white\,noise}$ (in mmag) were determined from a fit to the power spectrum by means of a Levenberg-Marquardt algorithm, and discarding the isolated dominant peaks in the periodogram from the fit. The errors on the parameters were estimated from the diagonal elements of the variance-covariance matrix \citep[see][]{Naze21}. The best-fit red noise parameters are listed in Table\,\ref{tab:periodogram} whilst Fig.\,\ref{rnfit} provides an illustration of the fit in the case of the periodogram of $\lambda$~Cep for sectors 16 and 17. The fits of the red noise were performed up to the Nyquist frequencies (25\,d$^{-1}$ for the 30\,min cadence data and 360\,d$^{-1}$ for the 2\,min cadence lightcurves). For HD~14\,442, the white noise level was not reached at 25\,d$^{-1}$, and we could only estimate an upper limit on $C_{\rm white\,noise}$. Therefore, the noise parameters of this star should be considered preliminary. The mean lifetime of the red noise features is comparable to that found for other OB-type stars displaying such variations in their lightcurve \citep{Mahy,Blomme,Rau19,Bow19,Bow20}. Overall, the values of $A_0$, $\tau$ and $\gamma$ nicely fit into the distribution of these parameters as a function of the spectroscopic luminosity\footnote{The spectroscopic luminosity, or inverse of the flux-mean gravity is defined as $\mathscr{L} = \frac{T_{\rm eff}^4}{g}$ \citep{LanKud}.} and effective temperature found for a sample of OB stars by \citet{Bow20}. For instance, the cooler $\lambda$~Cep and BD$+60^{\circ}$\,2522 display larger $A_0$ and $\tau$ than the three other, hotter, targets.

The presence of a red noise component also impacts the significance level of the various peaks listed in Table\,\ref{tab:periodogram}. To illustrate this point, let us again consider the case of the $\lambda$~Cep data of sectors 16 \& 17. The mean amplitude of the peak at $\nu_{{\rm TESS,}~\lambda~{\rm Cep}} = 0.414$\,d$^{-1}$ is $\sim 4.8$ times larger than the level of the (red + white) noise component at this frequency. To assess the significance of this peak, we first used the empirical formula of \citet[][their Eq.\,2]{Mahy} and \citet[][their Eq.\ 6]{Blomme}. Given the huge number of data points in the relevant lightcurve (30\,226), this formula leads to extremely low probabilities (consistent with zero) that the observed signal might result from a stochastic process. However, this formula applies to periodograms affected by white noise only, thus might not be appropriate in the present case, especially not at low frequencies.
\begin{figure}
\begin{center}
\resizebox{8cm}{!}{\includegraphics{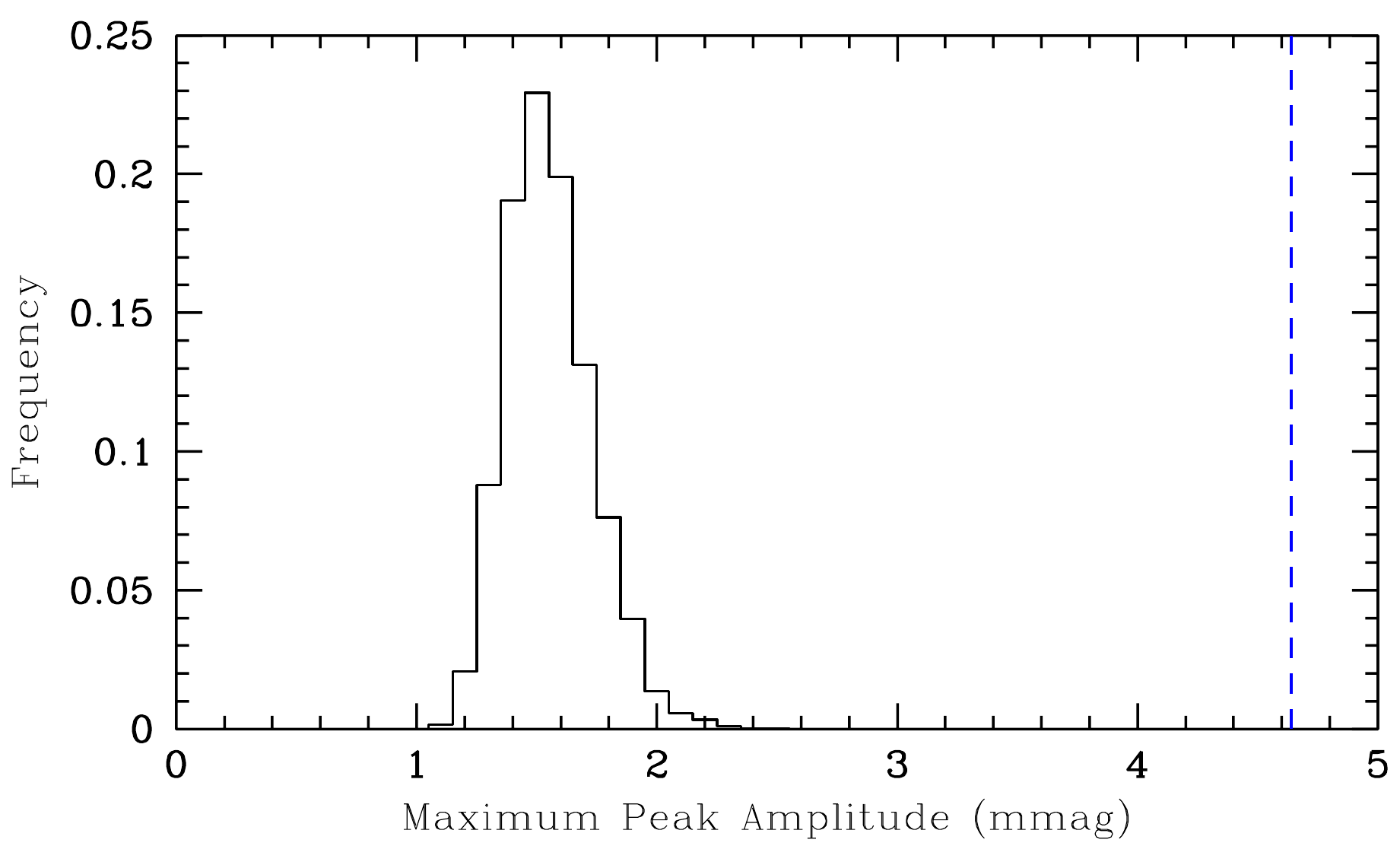}}
\end{center}
\begin{center}
\resizebox{8cm}{!}{\includegraphics{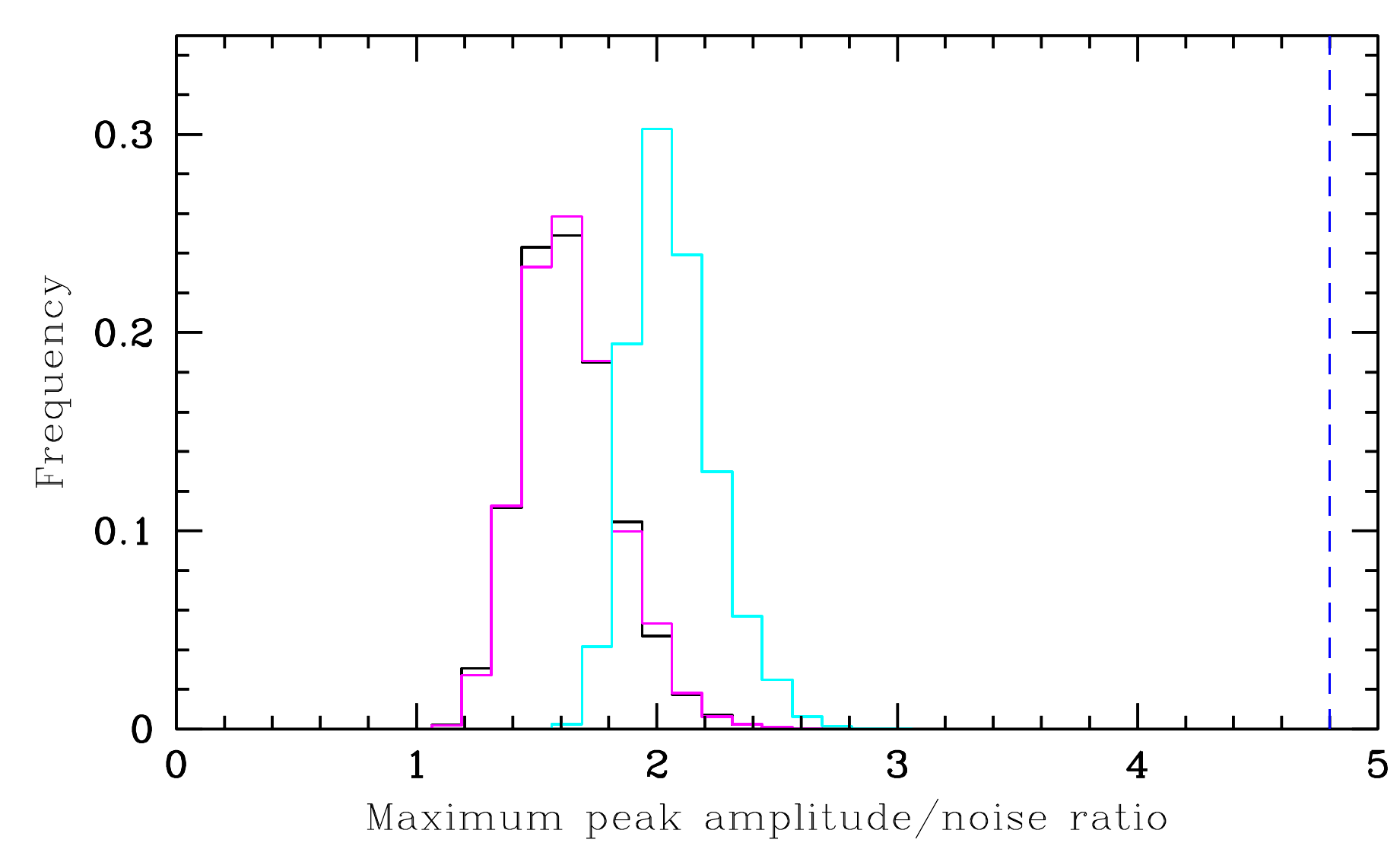}}
\end{center}
\caption{{\it Top:} Histogram of the maximum peak amplitude in the Fourier periodogram (computed up to 30\,d$^{-1}$) of the simulated lightcurves with the red and white noise parameters of $\lambda$~Cep for sectors 16 \& 17. The dashed blue vertical line indicates the amplitude of the $\nu_{{\rm TESS,}~\lambda~{\rm Cep}}$ peak derived from the full sector 16 \& 17 {\it TESS} PDC lightcurve of $\lambda$~Cep. {\it Bottom:} Histogram of the ratio between the peak amplitude and the noise level at the frequency of the highest peak in the Fourier periodogram of the synthetic lightcurves. The black, magenta and cyan histograms stand for the results obtained for frequencies below 1\,d$^{-1}$, between 1 and 2\,d$^{-1}$ and between 2 and 10\,d$^{-1}$, respectively. The dashed blue vertical line yields the ratio measured for $\nu_{{\rm TESS,}~\lambda~{\rm Cep}}$ on the periodogram of the observed data.\label{histogram}}
\end{figure}

We therefore generated 5\,000 synthetic lightcurves using a Monte Carlo simulator and adopting the same sampling as for the actual {\it TESS} PDC lightcurves. These simulations assume that the photometric variability is entirely due to noise following Eq.\,\ref{eq1}. The red noise was simulated using the recipe of \citet{TK} with the best-fit model parameters listed in Table \ref{tab:periodogram}. For each synthetic lightcurve, a Fourier periodogram was computed in exactly the same manner as for the actual data. Finally, histograms of the maximum peak amplitude of the periodogram as well as of the maximum value of the ratio between the peak amplitude and the amplitude of the (red + white) noise level at the corresponding frequency were built (see Fig.\,\ref{histogram}). As the red noise level changes with frequency, the distribution of the maximum amplitudes changes with frequency and even the amplitude/noise ratio of the synthetic data depends on the frequency domain that is considered in the periodogram. This method leads to a clear assessment of the detected signals (see third column of Table\,\ref{tab:periodogram}): for frequencies below 2\,d$^{-1}$, a peak of amplitude exceeding 2.5 times the local (red + white) noise level is unlikely to be due to a fluctuation of the red noise component. We thus find that significant peaks are detected in each target. This is also backed up by the fact that these strong peaks remain visible over a two orders of magnitude longer timescale than the mean lifetime of the red noise features (see Figs.\,\ref{spevol} and \ref{FigAppen1}, plus Table\,\ref{tab:periodogram}).

\subsection{Other photometric time series}
To complement the {\it TESS} data, we searched for additional photometric time series of our targets.
For $\lambda$~Cep, we re-analysed the {\it Hipparcos} photometry for which \citet{Mar98} previously reported the detection of a 0.63\,d modulation with an amplitude of 8\,mmag. We used the same Fourier method as for the {\it TESS} data. Discarding three data points with formal errors larger than 0.010\,mag, the full sample consists of 124 measurements spread over 1\,157\,days. The formal errors range between 3 and 8\,mmag with a mean value of 5\,mmag. The Fourier periodogram reveals no clear periodicity. Whilst the peak reported by \citet{Mar98} is present, it does not stand out against the noise and is thus not significant. There is no indication of other significant peaks. Similar conclusions apply to the {\it ASAS-SN}\footnote{https://asas-sn.osu.edu/ To avoid artefacts in the Fourier periodogram of these {\it ASAS-SN} data, we discarded isolated data points taken before JD~2\,456\,500 and removed outliers by means of a median absolute deviation filtering.} \citep{sha14} data of the targets, for which errors amount to 5--7\,mmag. The failure to detect any signal comparable to what was found with {\it TESS} can be explained by the typical errors on the {\it Hipparcos} and {\it ASAS-SN} data combined with the typical lifetime and amplitude of the signals seen in the {\it TESS} photometry.
An interesting result is a $0.034$\,d$^{-1}$ frequency (period of 29.4\,d) detected with an amplitude of 32\,mmag in the {\it ASAS-SN} data of $\lambda$~Cep. We stress however that no indication of this frequency was found in the {\it TESS} data of this star. 

\begin{figure}
\begin{center}
  \resizebox{9cm}{!}{\includegraphics{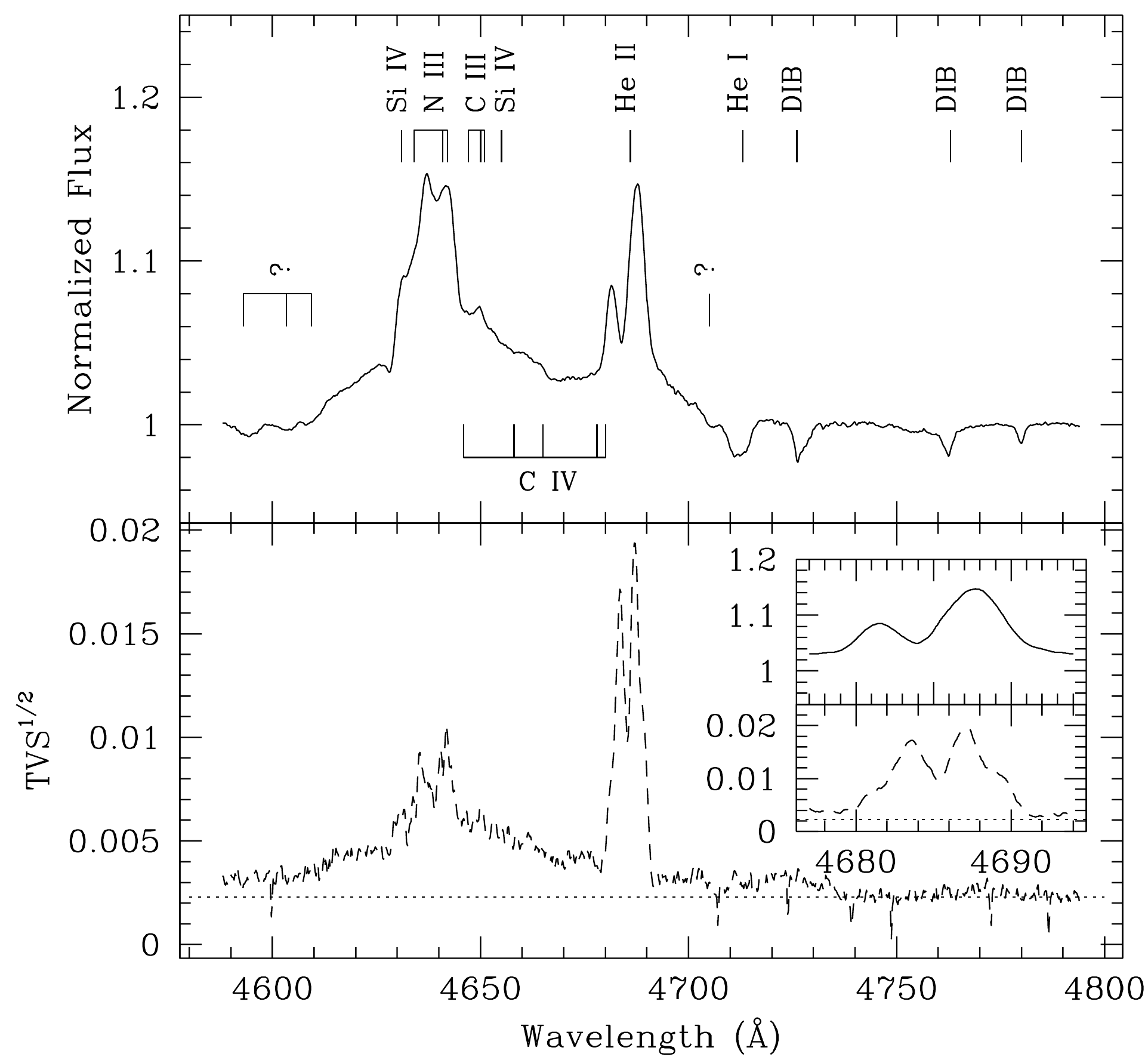}}
\end{center}  
\caption{Mean normalized spectrum and temporal variance spectrum (TVS) computed from the June 2015 Aur\'elie spectra of $\lambda$~Cep. The dotted line in the lower panel yields the 99\% significance level. The insert in the lower panel provides a zoom on the region of the He\,{\sc ii} $\lambda$\,4686 line.\label{meanspec}}
\end{figure}

\subsection{Spectroscopy of $\lambda$~Cep} \label{spectro}
As a complement to the {\it TESS} data of $\lambda$~Cep, we briefly discuss the spectroscopic variability of this star as observed during June 2015. The mean spectrum and the temporal variance spectrum \citep[TVS,][]{FGB} are shown in Fig.\,\ref{meanspec}. Some aspects of the mean spectrum are discussed in Appendix\,\ref{Appendix2}. Highly significant variability is found in the N\,{\sc iii} $\lambda\lambda$\,4634-4641 and He\,{\sc ii} $\lambda$\,4686 emission lines as well as in the broad emission bump between 4600 and 4700\,\AA. On the contrary, the level of variability is quite low in the He\,{\sc i} $\lambda$\,4713 line.

To characterize the spectral variability over a certain wavelength interval, the time series of Aur\'elie spectra was analysed by means of the 2D-Fourier analysis tool described by \citet{Rau08}. In this method, the Fourier periodogram is computed at each wavelength step between $\lambda_{\rm start}$ and $\lambda_{\rm end}$ using again the method of \citet{HMM} and \citet{Gosset} to account for the highly irregular temporal sampling. The wavelength-integrated power spectrum $P({\nu})$ is then computed as
\begin{equation}
  P({\nu}) = \sum_{\lambda_{\rm start}}^{\lambda_{\rm end}}\left(\frac{A_{\lambda}({\nu})}{2}\right)^2
  \label{power}
\end{equation}
where $A_{\lambda}({\nu})$ is the amplitude of the variations at frequency $\nu$ and for the spectral element of wavelength $\lambda$.
\begin{figure}
\begin{center}
  \resizebox{9cm}{!}{\includegraphics{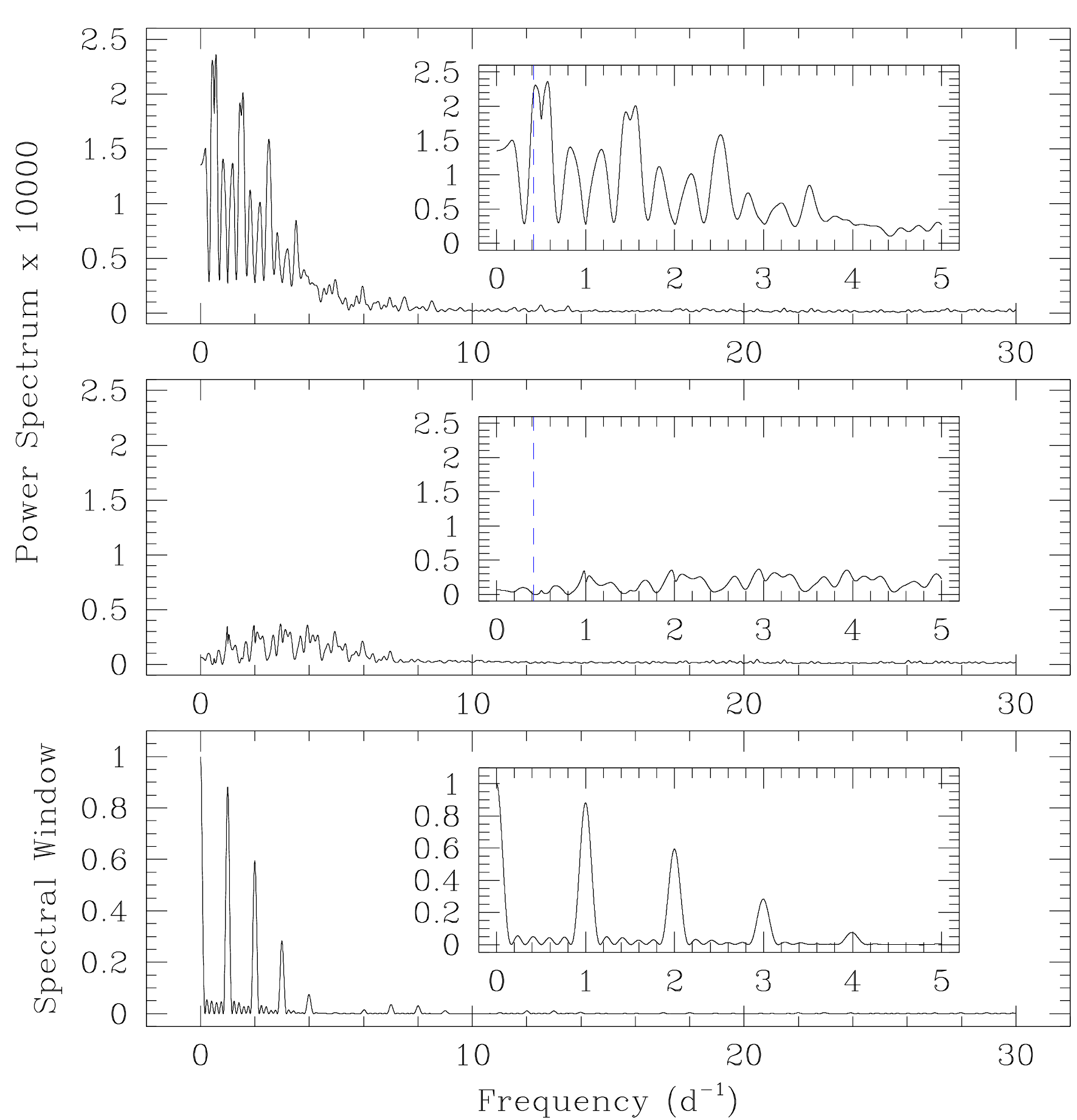}}
\end{center}  
\caption{Power spectrum of the variations of the He\,{\sc ii} $\lambda$\,4686 line integrated between 4677 and 4694\,\AA. From top to bottom, the different panels illustrate the power spectrum computed from the original data, the power spectrum after prewhitening of the frequencies 0.44\,d$^{-1}$ and 0.80\,d$^{-1}$, and the spectral window. The inserts provide a zoom on the frequency domain up to 5\,d$^{-1}$. The dashed blue line indicates the frequency $\nu_{{\rm TESS,}~\lambda~{\rm Cep}} = 0.414$\,d$^{-1}$ found in the photometric data from sectors 16 \& 17. \label{fourier4686}}
\end{figure}

The highest peaks in the Fourier periodogram  of the He\,{\sc ii} $\lambda$\,4686 line integrated between 4677 and 4694\,\AA\ are found at a frequency of 0.57\,d$^{-1}$ and its $1 - \nu$ alias 0.44\,d$^{-1}$, which has only marginally lower height (Fig.\,\ref{fourier4686}). Because of the natural widths of the peaks (0.19\,d$^{-1}$), these aliases are heavily blended and their maxima are therefore shifted towards 0.5\,d$^{-1}$.

To search for additional frequencies, the data were prewhitened. For this purpose, a relation of the kind
\begin{equation}
  y(t,\lambda) = a_{0}(\lambda) + \sum_{j=1}^N a_j(\lambda)\,\sin{[2\,\pi\nu_j\,t + \psi_j(\lambda)]}
  \label{relprew}
\end{equation}
is first adjusted to the data. Here $y(t,\lambda)$ stands for the time series of data at wavelength $\lambda$, $N$ is the number of frequencies $\nu_j$ to be considered. The parameters to be determined are the mean normalized spectrum $a_0(\lambda)$, the amplitudes $a_j(\lambda)$ and the phases $\psi_j(\lambda)$ of the sine waves. Once the best-fit parameters have been found, the signal at the adjusted frequencies is subtracted from the observed time series, and the Fourier analysis is repeated on the prewhitened time series.  

Prewhitening the He\,{\sc ii} $\lambda$\,4686 data for either of the two frequencies, 0.44 or 0.57\,d$^{-1}$ yields identical results, and leaves significant residual power in a series of peaks. The highest remaining peak is located at 0.20\,d$^{-1}$, and its alias at 0.80\,d$^{-1}$ has nearly identical amplitude. Prewhitening for various combinations of two frequencies, either 0.44 and 0.20\,d$^{-1}$, 0.57 and 0.20\,d$^{-1}$ or 0.44 and 0.80\,d$^{-1}$ removes most of the power (see Fig.\,\ref{fourier4686}).

Since the 0.44\,d$^{-1}$ and 0.80\,d$^{-1}$ frequencies are relatively close respectively to the $0.414$\,d$^{-1}$ frequency found in the {\it TESS} data from sectors 16 and 17 (see Sect.\,\ref{TESSanalyse}) and twice this frequency, we have also prewhitened the He\,{\sc ii} $\lambda$\,4686 time series for $\nu_{{\rm TESS,}~\lambda~{\rm Cep}}$ and $2\,\nu_{{\rm TESS,}~\lambda~{\rm Cep}}$. The results are nearly indistinguishable from those obtained with the 0.44\,d$^{-1}$ and 0.80\,d$^{-1}$ frequencies. This result indicates that the variations of the He\,{\sc ii} $\lambda$\,4686 line can in principle be explained by $\nu_{{\rm TESS,}~\lambda~{\rm Cep}}$ and $2\,\nu_{{\rm TESS,}~\lambda~{\rm Cep}}$. 

\begin{figure}
\begin{center}
  \resizebox{9cm}{!}{\includegraphics{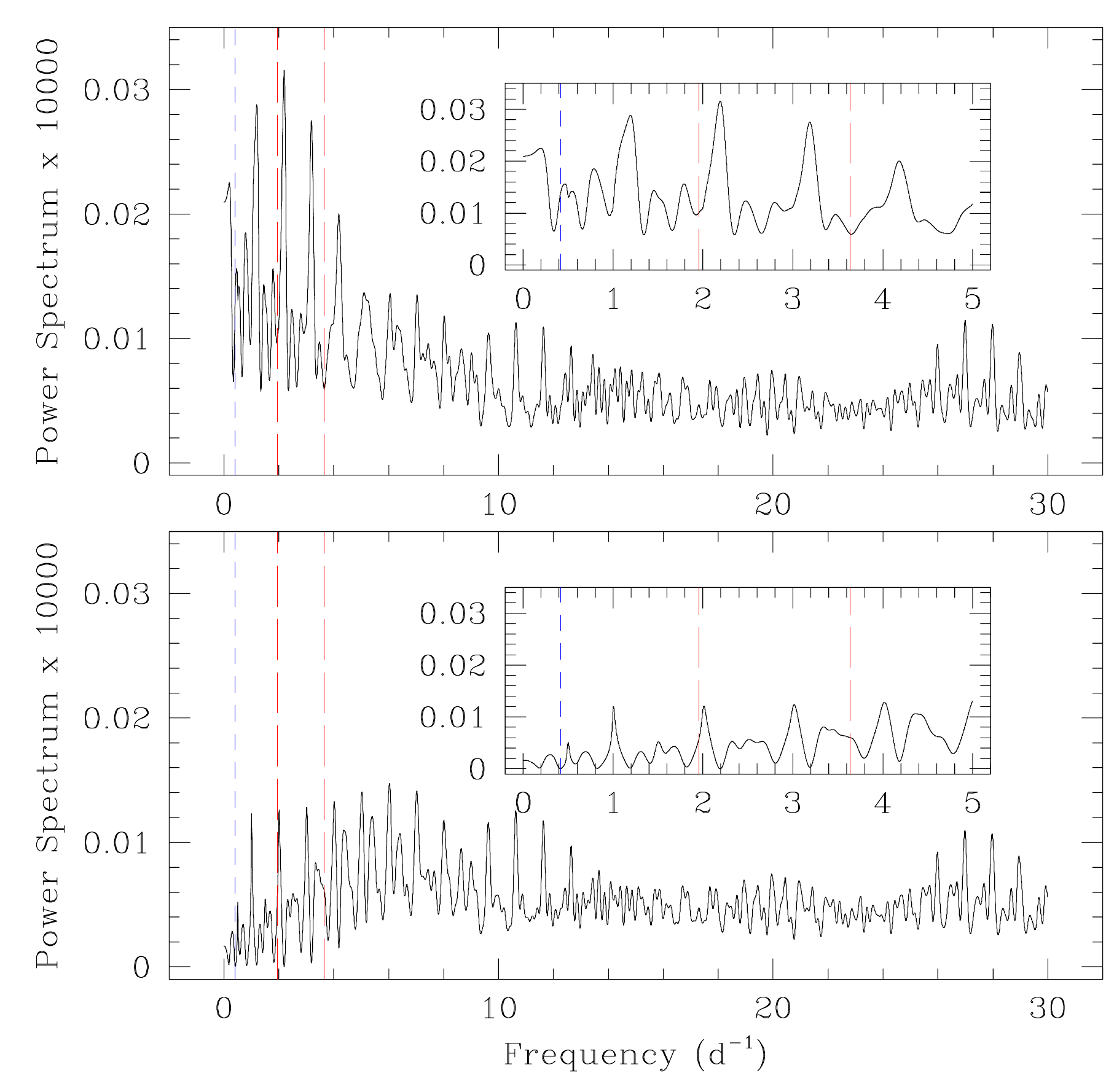}}
\end{center}  
\caption{{\it Top panel:} Fourier power spectrum of the variations of the He\,{\sc i} $\lambda$\,4713 line integrated between 4706 and 4718\,\AA. {\it Bottom panel:} power spectrum after prewhitening for the frequencies 2.19\,d$^{-1}$, $\nu_{{\rm TESS,}~\lambda~{\rm Cep}}$ and $2\,\nu_{{\rm TESS,}~\lambda~{\rm Cep}}$. The short-dashed blue line corresponds to $\nu_{{\rm TESS,}~\lambda~{\rm Cep}}$, whereas the two long-dashed red lines indicate the frequencies previously reported by \citet{deJ99}.\label{fourier4713}}
\end{figure}

An interesting question is whether or not the photospheric absorption line He\,{\sc i} $\lambda$\,4713 displays the signatures of non-radial pulsations. The TVS$^{1/2}$ of this line exceeds the 99\% significance level, although only by a small amount. The low-frequency part of the Fourier power spectrum is dominated by a peak at 2.19\,d$^{-1}$ (see Fig.\,\ref{fourier4713}). Because of the aliasing, there certainly exists no unique combination of frequencies that can account for the full content of the power spectrum. Most of the power can be accounted for by the 2.19\,d$^{-1}$ frequency combined again with $\nu_{{\rm TESS,}~\lambda~{\rm Cep}}$ and $2\,\nu_{{\rm TESS,}~\lambda~{\rm Cep}}$. After prewhitening, we find a residual peak near the 1.96\,d$^{-1}$ frequency (see Fig.\,\ref{fourier4713}) that was previously associated to NRPs by \citet{deJ99}. However, we note that this peak is at the level of the noise found e.g.\ in the power spectra of the neighbouring diffuse interstellar bands at 4726 and 4762\,\AA\ and that the properties of this signal (amplitude and phase) depend strongly on the choice of the frequencies that are included in the prewhitening process. 

As a conclusion, the multi-epoch data of $\lambda$~Cep allowed us to show that the discrete frequency at 0.414\,d$^{-1}$, which dominated the {\it TESS} observations in autumn 2019, was also possibly present in the spectroscopic variability in June 2015 but had disappeared from the photometric variations by May 2020.

\section{Discussion}\label{discussion}
\subsection{Revised properties of the Onfp/Oef stars}\label{revision}
Before we turn to the interpretation of the observed variability, it is worth coming back to some properties of the stars in our sample.  
\begin{table*}
  \caption{Inferred properties of our target stars}
  \begin{tabular}{c c c c c c c}
    \hline
    Star & $M_V$ & Lumin.\ class & $R/R_{\odot}$ & $\nu_{\rm rot}^{\rm min}$ & $\nu_{\rm rot}^{\rm max}$ & $\log{{\mathscr{L}}/{\mathscr{L}_{\odot}}}$  \\
         &       &               &             & (d$^{-1}$)  & (d$^{-1}$)   &  \\
    \hline\hline
    $\lambda$~Cep & $-5.48 \pm 0.17$ & III & $14.0 \pm 1.8$ & $0.30 \pm 0.04$ & $0.81 \pm 0.16$ &  $4.06 \pm 0.17$ \\
    HD~14\,434    & $-4.71 \pm 0.24$ &  V  & $9.2 \pm 1.3$  & $0.87 \pm 0.13$ & $1.61 \pm 0.35$ & $3.77 \pm 0.16$  \\
    HD~14\,442    & $-5.58 \pm 0.27$ & III & $13.9 \pm 2.1$ & $0.40 \pm 0.06$ & $0.77 \pm 0.18$ & $3.98 \pm 0.16$ \\
    HD~192\,281   & $-4.97 \pm 0.11$ &  V  & $10.6 \pm 1.1$ & $0.51 \pm 0.06$ & $1.28 \pm 0.21$ & $4.03 \pm 0.16$ \\
    BD$+60^{\circ}$\,2522 & $-5.40 \pm 0.18$ & III & $13.3 \pm 2.1$ & $0.35 \pm 0.06$  & $0.88 \pm 0.21$ & $4.08 \pm 0.15$\\
    \hline
    HD~93\,521    & $-4.30 \pm 0.49$ &  V  & $9.2 \pm 2.2$ & $0.87 \pm 0.21$ & & $3.52 \pm 0.12$  \\
    \hline
  \end{tabular}
  \label{keyprop2}
  
  {\scriptsize }  
\end {table*}
Based on the magnitudes of Onfp/Oef stars in the LMC and SMC, \citet{Wal10} pointed out that the Onfp phenomenon occurs at all luminosities, not just for giants and supergiants. These authors therefore concluded that conventional luminosity criteria for O-stars, based on the nature of the He\,{\sc ii} $\lambda$\,4686 line, do not work for Onfp/Oef stars. We have used the magnitudes, colours and {\it GAIA}-DR2 parallaxes listed in Table\,\ref{keyprop} to compute absolute magnitudes of our target stars. For this purpose, we adopted an intrinsic colour $(B-V)_0 = -0.27$ \citep{MP} and $R_V = 3.1$. We further added an offset of 0.03\,mas to the {\it GAIA}-DR2 parallax \citep{Lindegren}. The results are listed in Table\,\ref{keyprop2} along with the luminosity classes that best match the `typical' absolute magnitudes\footnote{Given its location far away from the Galactic plane, there has been some discussion about the nature of HD~93\,521 \citep[see][and references therein]{Rau12}. However, the {\it GAIA}-DR2 parallax of HD~93\,521 rules out the possibility that it could be a hot subdwarf and the inferred $M_V$ clearly indicates a genuine Population I O-star.} of \citet{Martins}. We find that all our Onfp/Oef sample stars are consistent with main-sequence or giant stars, thereby confirming the assertion of \citet{Wal10} that the nature and strength of the He\,{\sc ii} $\lambda$\,4686 line of these stars cannot be used as a criterion of the luminosity class. An interesting case is $\lambda$~Cep which is usually classified as a supergiant. Whilst {\it GAIA}-DR2 parallaxes of stars brighter than sixth magnitude must be considered with caution, we note that the {\it GAIA} parallax of $\lambda$~Cep was determined at the $12.7\,\sigma$ level and indicates a distance around 0.61\,kpc \citep{Shull}. This is significantly lower than the usually adopted distance of 0.7 -- 1.0\,kpc \citep[see][]{SudHen16,Shull}. Yet, the values of $\log{g} = 3.50 \pm 0.15$, $3.39 \pm 0.06$ or $3.55 \pm 0.10$ inferred by \citet{Mar15,Cazorla}, \citet{Hol20} and \citet{Rep04} from model atmosphere fitting favour a supergiant luminosity class \citep[$\log{g} = 3.48$ for an O6\,I,][]{Martins}, although the giant luminosity class \citep[$\log{g} = 3.65$ for an O6\,III star,][]{Martins} lies within the errors of most of the above $\log{g}$ values. Assuming $\lambda$~Cep to be a supergiant would imply that the {\it GAIA}-DR2 parallax must be off by $3.5\,\sigma$. It will be interesting to see whether forthcoming {\it GAIA} data releases confirm the lower value of the star's distance.

The observed high rotational velocities are easier to reconcile with the revised luminosity classes. Indeed, during the evolution of a single massive star, the loss of angular momentum through the stellar winds is expected to slow down the rotation. If Onfp/Oef stars were indeed evolved supergiants, their rotation rates would be extremely unusual \citep{Wal10}, unless they were significantly spun-up by angular momentum transfer from a companion that fills up its Roche lobe\footnote{In this context, we note that whilst no evidence for binarity has been found for any of the Onfp/Oef stars of our sample \citep{DeB04,Rau03,Rau15}, $\lambda$~Cep is considered a runaway star \citep{Gies}. The current properties of this star could thus be altered by a mass- and angular momentum exchange episode in a now-disrupted binary system.}. The downwards revised luminosities partially solve this issue. At this stage, it is worth pointing out that the CNO abundances inferred by \citet{Cazorla} for our sample Onfp/Oef stars indicate mildly-evolved stars showing a moderate chemical enrichment consistent with the predictions of stellar evolution models accounting for rotational mixing \citep[e.g.][]{Eks12}.  

With the absolute magnitudes at hand, we have then used the bolometric corrections computed according to \citet{LanHub03} to establish the bolometric magnitudes, and from there, with the effective temperatures listed in Table\,\ref{specprop} the stellar radii (Table\,\ref{keyprop2}). These radii and the observed $v\,\sin{i}$ were used to infer the minimum rotational frequencies (see Table\,\ref{keyprop2}). Assuming a stellar mass between 30 and 35\,M$_{\odot}$, typical for mid O\,V-III stars \citep{Martins}, for the Onfp/Oef stars, we have also computed the upper limit on the rotational frequency corresponding to critical equatorial rotation. Finally, the last column of Table\,\ref{keyprop2} lists the spectroscopic luminosities based on the effective temperatures and surface gravities corrected for centrifugal forces taken from \citet{Cazorla} (except for BD$+60^{\circ}$\,2522 for which the values are from \citealt{Hol20}) quoted in Table\,\ref{specprop}.

\subsection{Comparison with $\zeta$~Pup}\label{compazetaPup}
Our present study revealed that the photometric variability of Onfp/Oef stars consists of a strong red noise component associated with a single discrete frequency which however does not constitute a persistent feature. This is similar to the situation of the spectroscopic variability. For instance, as clearly demonstrated by \citet{Uuh14}, \citet{Rau15}, \citet{SudHen16}, and the present work, the line profile variability of $\lambda$~Cep is not ruled by a single stable period. Indeed, \citet{SudHen16} list about a dozen different periods that show up at different epochs in the line profile variability of UV or optical features. Likewise, the He\,{\sc ii} $\lambda$\,4686 line of BD$+60^{\circ}$\,2522 varies on timescales of 2 to 3 days \citep{Rau03}, but the pattern of variability is epoch-dependent and observing campaigns separated by just one month yielded different periodograms \citep{Rau03}. A similar situation holds for the spectroscopic variability of the other Onfp/Oef stars of our sample \citep{DeB04}.

At this stage, it is interesting to compare our sample to the brightest Onfp/Oef object, the O4\,Ief star $\zeta$~Pup. \citet{Tahina} investigated photometric data from a 5.5\,month {\it BRITE} campaign of $\zeta$~Pup. Besides a red noise component that accounts for a significant fraction of the photometric variability of $\zeta$~Pup, they confirmed the presence of a 1.78\,d periodicity which had been previously reported from {\it SMEI} photometry \citep{2Ian}. \citet{Tahina} further found the modulation to change its shape with time and to be highly non-sinuoidal. On average, the fundamental frequency had a mean amplitude of $\sim 3.8$\,mmag, whilst the second harmonic had a mean amplitude of $\sim 2.5$\,mmag. Whilst \citet{2Ian} advocated a pulsational origin for the signal, \citet{Tahina} argue that the highly non-sinusoidal shape and its time-dependence imply that the modulation of $\zeta$~Pup rather comes from rotational modulation due to spots appearing and disappearing at the stellar surface.

The time-frequency diagrams of our Onfp/Oef stars also shows red noise and isolated peaks but no harmonic is detected and, furthermore, a major difference between our stars and $\zeta$~Pup concerns the long-term stability of these peaks. The 1.78\,d signal in $\zeta$~Pup seems to affect the photometric variability of the star over timescales of many years, probably even decades while our analysis rather shows transient features, see notably the spectacular example of the $0.414$\,d$^{-1}$ frequency in $\lambda$~Cep.

\subsection{Rotational modulation?}
The dominant frequencies reported in Table\,\ref{periodogram} fall in the range between the minimum and maximum rotational frequencies given in Table\,\ref{keyprop2}, with the notable exceptions of the $1.230$\,d$^{-1}$ frequency observed for HD~14\,442 and the $0.194$\,d$^{-1}$ frequency found in the sector 24 data of BD$+60^{\circ}$\,2522. In this section, we thus consider the pros and cons of an interpretation of the dominant frequency as the rotational frequency.

To explain the spectroscopic variability of $\lambda$~Cep, \citet{Hen14} and \citet{SudHen16} proposed the existence of so-called stellar prominences associated with short-lived corotating magnetic loops. Variability would then arise from the changing viewing angle as the star rotates \citep{SudHen16}. \citet{Gru17} presented spectro-polarimetric observations of $\lambda$~Cep and HD~192\,281. No global dipolar magnetic field was detected. This is consistent with the upper limit on a dipolar magnetic field of 136\,G inferred by \citet{DU14} for $\lambda$~Cep. Whilst strong global dipolar magnetic fields seem unlikely, the scenario of \citet{SudHen16} requires only localized magnetic spots that might escape detection with current instrumentation. In massive stars, localized magnetic fields (of strength a few hundred G) can be produced by a convective zone close to the stellar surface which stems from a peak in opacity associated with the iron group elements \citep{CB11}. These fields could emerge at the stellar surface via magnetic buoyancy, thereby producing hot ($\Delta T$ of a few hundred K) and bright spots. This scenario could account for two properties of the dominant frequencies that we have found. Indeed, the fact that the dominant frequencies can disappear (as highlighted by the non-detection of $\nu_{{\rm TESS,}~\lambda~{\rm Cep}}$ in the $\lambda$~Cep sector 24 data) clearly indicates that the features responsible for these modulations must be transient as one would expect for a magnetic spot in the \citet{CB11} scenario. Moreover, the slight variations of the dominant frequencies on timescales of tens of days (see Figs.\,\ref{spevol} and \ref{FigAppen1}) could be the result of a latitude-dependent rotational velocity if the spots responsible for the modulation drift in stellar latitude. 

Such bright spots could trigger CIRs in the stellar wind as shown by \citet{DaUr17}. Based on a time series of {\it IUE} spectra, \citet{MP15} indeed reported the signature of CIRs in the Si\,{\sc iv} $\lambda\lambda$\,1400 resonance doublet and in the excited state wind line N\,{\sc iv} $\lambda$\,1718 of $\lambda$~Cep. These authors noted that the features responsible for the CIRs should cause a 2 -- 4\% modulation in the photospheric flux. For our Onfp/Oef targets, the {\it TESS} data indicate peak-to-peak variability at about the right level. However, most of this photometric variability is very likely stochastic and the amplitude of the dominant frequencies, possibly associated with rotation, is rather of the order 5\,mmag. This is quite comparable with the situation of $\zeta$~Pup and of the O7.5\,III(n)((f)) star $\xi$~Per which also have rather low photometric variability despite these stars exhibiting clear signatures of CIRs in their UV spectra \citep{MP15,mas19}. In this context, \citet{Rau15} observed that the X-ray flux variations of $\lambda$~Cep were nearly anti-correlated with the H$\alpha$ equivalent width variations. This anti-correlation was explained in the context of the magnetic spots model as resulting from a magnetically-heated highly-ionized plasma produced by the spot at the base of the CIR. This hot plasma would create a hole in the H$\alpha$ emitting region above the spot. However, this correlation might as well be a coincidence. In fact, in $\zeta$~Pup, the X-ray flux appears to display a variety of behaviours, with only occasionally a direct correlation with the 1.78d period \citep{NRS18}. Further X-ray investigation confirmed the presence of the 1.78d signal, but with a different shape than the optical lightcurve and a shift between the X-ray and optical maxima (Nichols et al. 2020, submitted). Similarly, $\xi$~Per shows a shift between optical spectroscopic changes and X-ray variations \citep{mas19}. 

A number of frequencies have been reported previously in the spectroscopic variability studies of our Onfp/Oef sample stars (see Tables 4 and 5 of \citealt{SudHen16} for $\lambda$~Cep, \citealt{Rau03} for BD$+60^{\circ}$\,2522 and \citealt{DeB04} for the other three stars). Most of these frequencies are not related in a simple way to the dominant frequencies found in the {\it TESS} photometry. The same limitation holds for the $0.245$\,d$^{-1}$ frequency that \citet{SudHen16} proposed to be the rotational frequency of $\lambda$~Cep, although this frequency falls outside the range of possible rotational frequencies in Table\,\ref{keyprop2}. In their model, the wealth of observed frequencies then results from the interplay of the lifetimes of the stellar prominences and the rotational modulation of their visibility. The best-fit lifetimes of the prominences range from a few hours up to one day \citep{SudHen16} and are considerably shorter than the lifetimes of the features responsible for the dominant frequencies in the {\it TESS} data but much longer than the lifetime of red noise features (Table\,\ref{tab:periodogram}). The limitations of ground-based spectroscopic campaigns (aliasing, day/night gaps, etc.) could possibly help explain some of the frequencies reported by \citet{SudHen16} without having to postulate very short lifetimes of the responsible transient features. Nevertheless, it seems unlikely that all frequencies can be traced to a single (rotational) frequency in Onfp/Oef stars. Beside the interplay of other phenomena (such as the stochastic waves responsible for the red noise), an alternative for explaining the range of observed frequencies would be to assume that these stars have a rather strong dependence of their angular rotation velocity on stellar latitude. Surface differential rotation is known for our Sun and could also explain some features seen in late-type spotted rotators \citep{Rei15}, but no constraints are available for early-type stars such as our Onfp/Oef targets.

However, the frequencies observed for HD~14\,442 and BD$+60^{\circ}$\,2522 challenge the rotational scenario. In the case of BD$+60^{\circ}$\,2522, the problematic frequency does not stand out high above the red noise level and could thus simply arise from the same phenomenon as the red noise. The situation of HD~14\,442 is far more problematic as we are dealing with the strongest single peak found in any of our stars. Interestingly, all the spectroscopic frequencies reported by \citet{DeB04} fell into the range of possible $\nu_{\rm rot}$, whilst our photometric frequency is well above the break-up rotational frequency. Moreover, there is no evidence for binarity of this star \citep{DeB04}, so that the observed frequency is unlikely to come from a companion star. A possible solution would be to assume that the 1.230\,d$^{-1}$ dominant frequency in the {\it TESS} data actually corresponds to the second harmonic of $\nu_{\rm rot}$. However, given the strength of the dominant peak, it is strange that in this case, no significant signal is observed at the fundamental frequency (0.615\,d$^{-1}$). Indeed, having two identical spots on opposite sides of the star, as would be required to produce a sinusoidal signal at the second harmonic of $\nu_{\rm rot}$, seems rather unlikely. Actually, a similar problem exists for $\zeta$~Pup. Indeed, \citet{how19} demonstrated that a rotation at the photometric period of 1.78d would imply the star to rotate near criticality, an unlikely occurrence. Moreover, it would then be seen close to pole-on and such a configuration makes rotational variability difficult to be seen. Finally, these authors further showed that it is difficult to reconcile the DACs timescale with the photometric period.

\subsection{Non-radial pulsations?} \label{disNRP}
For the sake of comparison with the results obtained for the Onfp/Oef stars, let us first consider the case of HD~93\,521, a rapidly rotating O9.5\,Vp star known to display non-radial pulsations. The location of HD~93\,521 in the spectroscopic Hertzsprung-Russell diagram \citep[sHRD, $\log{T_{\rm eff}}=4.48$ -- $4.49$, $\log{{\cal L}/{\cal L_{\odot}}} = 3.52 \pm 0.12$;][]{Rau12,Cazorla} places this star close to the instability strip where both low-frequency $g$ and high-frequency $p$ modes are expected to co-exist. HD~93\,521 also lies at the lower luminosity border of the domain where strange mode oscillations are expected \citep{Godart}.

Line profile variations due to NRPs were first reported in the spectrum of HD~93\,521 by \citet{FGB91} and \citet{HR93}. Subsequent observations by \citet{How98} revealed three frequencies $\nu_1 = 13.68$\,d$^{-1}$, $\nu_2 = 8.31$\,d$^{-1}$ and $\nu_3 = 2.66$\,d$^{-1}$. A multi-epoch spectroscopic monitoring of \citet{Rau08} revealed that $\nu_2$ was visible at all epochs of observations, whilst $\nu_1$ disappeared from time to time, and $\nu_3$ was not seen at all in these data. 

The main result that comes out of our analysis of the {\it TESS} lightcurve of HD~93\,521 is the fact that the photometric amplitudes of the spectroscopic NRP modes ($\nu_1$ and $\nu_2$) are quite modest. As expected, the mode with the higher $\ell$ value ($\nu_1$) has a less pronounced photometric counterpart. Whilst there are some variations of the amplitude of these modes with time (especially for $\nu_2$), we observe no drifts of their frequencies (unlike what is seen for $\lambda$~Cep). Our results for HD~93\,521 are reminiscent of those of $\zeta$~Oph, another rapidly rotating O9.5\,V star displaying NRPs for which space-borne photometry revealed frequencies in addition to those found in ground-based spectroscopy. In addition, all the photometric modulations undergo large variations of their amplitudes on timescales of hundreds of days \citep{How14}. 

From a theoretical point of view, the Onfp/Oef stars populate a region of the sHRD where low $\ell$ NRPs due to the $\kappa$ mechanism in the iron opacity bump are expected \citep{Godart}. Such pulsations can take the form of $\beta$~Cep-like $p$-modes with frequencies in the range between 1 and 2\,d$^{-1}$, $g$-modes with a frequency range from 0.3 to 0.6\,d$^{-1}$, or strange modes with periods of a few hours \citep{Godart}. The majority of the theoretically excited modes are expected to have $\ell \leq 4$, which should make them detectable in high-precision photometry. 

From the observational point of view, there is currently no consensus on the presence or absence of NRPs in Onfp/Oef stars. The presence of NRPs in $\lambda$~Cep was claimed by \citet{deJ99} based on their analysis of a series of observations of the He\,{\sc i} $\lambda$\,4713 line collected from two sites over five consecutive nights. They reported two periodicities, a dominant one with $\nu = 1.96$\,d$^{-1}$ ($P = 12.3$\,h) and a second one with $\nu = 3.64$\,d$^{-1}$ ($P = 6.6$\,h), and interpreted these signals as NRPs with $\ell = 3$ and $\ell = 5$, respectively. The first frequency appeared stronger in the blue part of the line and \citet{deJ99} attributed this to the blend with the unidentified line on the blue side of the He\,{\sc i} line (see Appendix\,\ref{Appendix2}). The amplitude of the first mode was at most 0.0015 of the normalized continuum, whilst it reached 0.0010 for the second mode. \citet{Uuh14} presented a multi-epoch spectroscopic monitoring of the He\,{\sc i} $\lambda$\,4471 and He\,{\sc ii} $\lambda$\,4542 lines of $\lambda$~Cep which revealed that the frequency content of the periodogram of these lines considerably varies from epoch to epoch. Only a subset of the detected `periods' displayed the progressive variation of the phase across the line profile expected for NRPs. Moreover, no strictly stable signal was found. Instead of genuine persistent NRPs, \citet{Uuh14} detected the presence of a red noise component in the power spectrum of the time series of the He\,{\sc i} $\lambda$\,4471 line. \citet{SudHen16} challenged the conclusion of \citet{Uuh14,Uuh15}, arguing that the strong lines investigated by \citet{Uuh14} are contaminated by the wind, inhibiting the detection of NRPs, whereas the weaker He\,{\sc i} $\lambda$\,4713 should arise from much deeper in the photosphere and thus be free of such contamination. Thanks to our new spectra, we now show that the signals reported by \citet{deJ99} were indeed also transient, as for other spectral lines. In their observations of BD$+60^{\circ}$\,2522, \citet{Rau03} noted variations of the absorption lines on timescales of 4 -- 5 hours, which they tentatively suggested to be due to NRPs, although the sampling of the data was not optimal to study such short timescales. A dedicated intensive spectroscopic monitoring of HD~14\,434 revealed no variations on timescales of a few hours, attributable to NRPs \citep{DeB04}.

The periodograms of the {\it TESS} photometry of our Onfp/Oef targets (Fig.\,\ref{periodogram}) reveal no outstanding peaks in the frequency range typically covered by $\beta$~Cep stars \citep[between 3 and 10\,d$^{-1}$,][]{PP,Han19} or $\zeta$\,Oph variables \citep[][see also the case of HD~93\,521 above]{How14}. Neither the {\it TESS} lightcurves nor the spectroscopic data of the He\,{\sc i} $\lambda$~4713 line of $\lambda$~Cep presented in this paper thus provide solid evidence for the presence of persistent $p$-mode NRPs in these stars. 

The red noise seen in the periodograms of OB stars, and prominent in our data, has been attributed to randomly excited gravity waves in the convective core \citep{Bow20}. Although our Monte Carlo simulations suggest that it is unlikely that the dominant peaks in the periodograms correspond to fluctuations of the red noise, the possibility remains that these dominant peaks might reflect a strong but temporary excitation of a coherent $g$-mode standing wave pulsation. How exactly these modes would be excited is an open question, though. It is nevertheless interesting to note that, whilst the pulsation frequencies in HD~93\,521 are stable, their amplitudes undergo changes which shows that even for a confirmed persistent pulsator, the excitation mechanism is clearly a dynamical phenonmenon. 

\section{Conclusions}\label{conclusion}
In this paper, we analysed photometric data of a sample of Galactic Onfp/Oef stars to search for the signatures of variability with stable periods on the long and medium term, as had been found for the prototype of this class, $\zeta$~Pup. Our results, however, draw another picture. The photometric variability of our targets does not exhibit a periodicity that is stable over long timescales. Most of the photometric variability of stars in our sample is actually due to stochastic red noise. On top of this red noise component, our targets display a dominant signal that remains visible over timescales of weeks to months. For example, such a peak appeared strong for $\lambda$~Cep during at least two months in the autumn 2019, but had totally disappeared six months later. Whilst rotational modulation by a transient spot at the stellar surface could account for some of the dominant frequencies, the frequency found for HD~14\,442 is superior to the critical rotation frequency and challenges the rotational modulation scenario. The photometric data also strongly argue against the presence of persistent and stable $p$-mode non-radial pulsations. This conclusion is backed-up by spectroscopic data (notably a new campaign on $\lambda$~Cep) which also fail to reveal a clear signature of such variations. The pulsation-like variations that were previously reported were thus most likely transient features. Yet, the dominant frequencies seen in our periodograms might be associated with temporarily excited $g$-modes, though it is currently unclear what mechanism would trigger their excitation.   

\section*{Acknowledgements}
GR would like to express his gratitude to the technical staff of the Haute Provence Observatory, especially to the telescope operators J.-P.\ Bretagne, R.\ Giraud, D.\ Gravallon, J.-C.\ Mevolhon, and J.-P.\ Troncin. Their efficient help contributed a lot to the success of many observing nights. The authors thank Drs.\ A.\ Pigulski and J.-C.\ Bouret for discussion. Support from the Fonds National de la Recherche Scientifique (Belgium), the Communaut\'e Fran\c caise de Belgique (including notably support for the observing runs at OHP), and the Belgian Federal Science Policy Office (BELSPO) in the framework of the PRODEX Programme (contract HERMeS) is acknowledged. This OefTI (Oef {\it TESS} Investigation) project makes use of data collected by the TESS mission, whose funding is provided by the NASA Explorer Program. ADS and CDS were used for this research. We thank the referee, Dr.\ Sergio Sim\'on-D\'{\i}az for his report that helped us improve the presentation of our results.

\section*{Data availability}
The {\it TESS} data underlying this article are available from the MAST archives, while the OHP spectra are made available on the journal website.

\appendix
\section{Time-frequency diagrams} \label{Appendix1}
\begin{figure*}
\begin{minipage}{8.5cm}
  \begin{center}
    \resizebox{!}{5.5cm}{\includegraphics{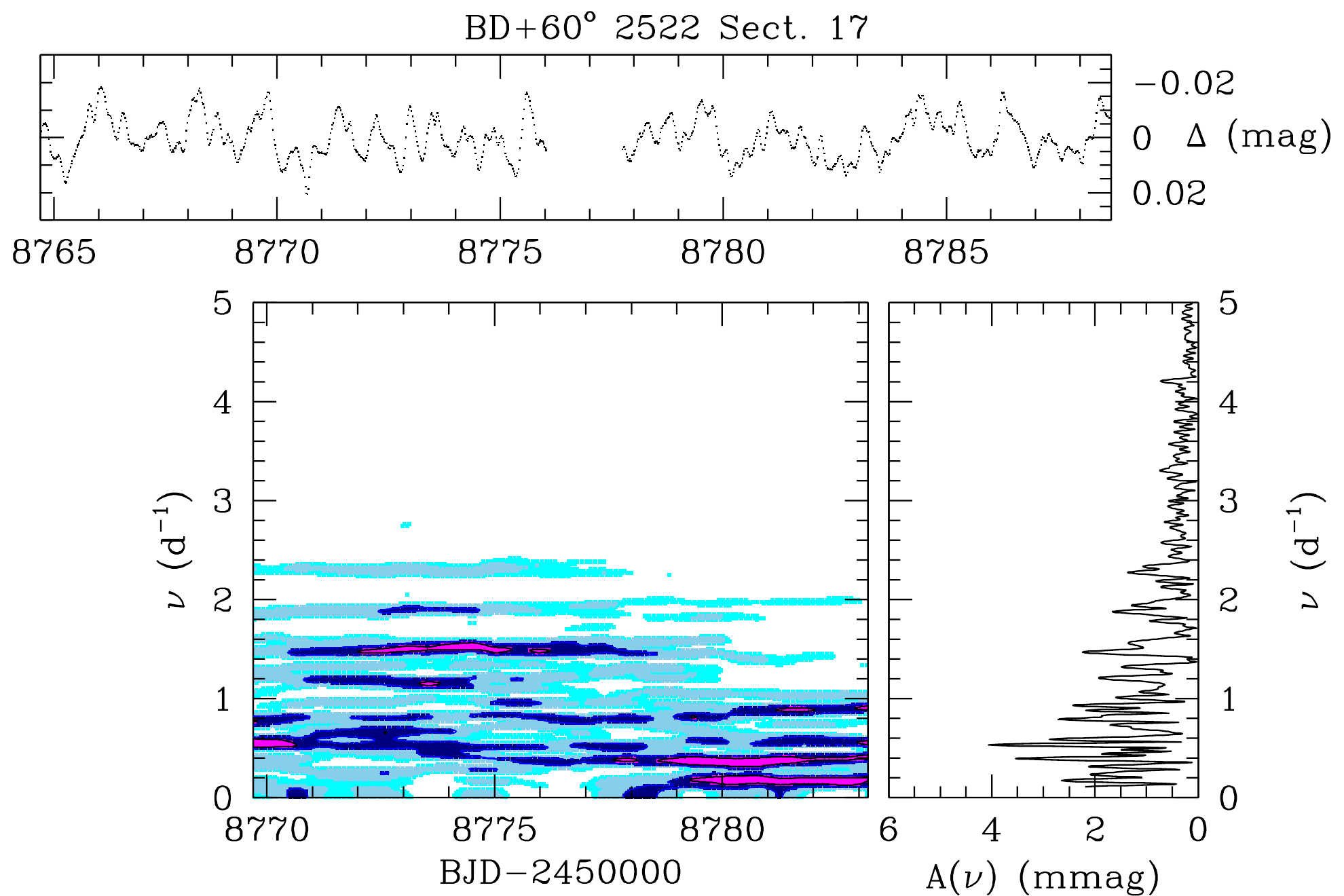}}
  \end{center}
\end{minipage}
\begin{minipage}{8.5cm}
  \begin{center}
    \resizebox{!}{5.5cm}{\includegraphics{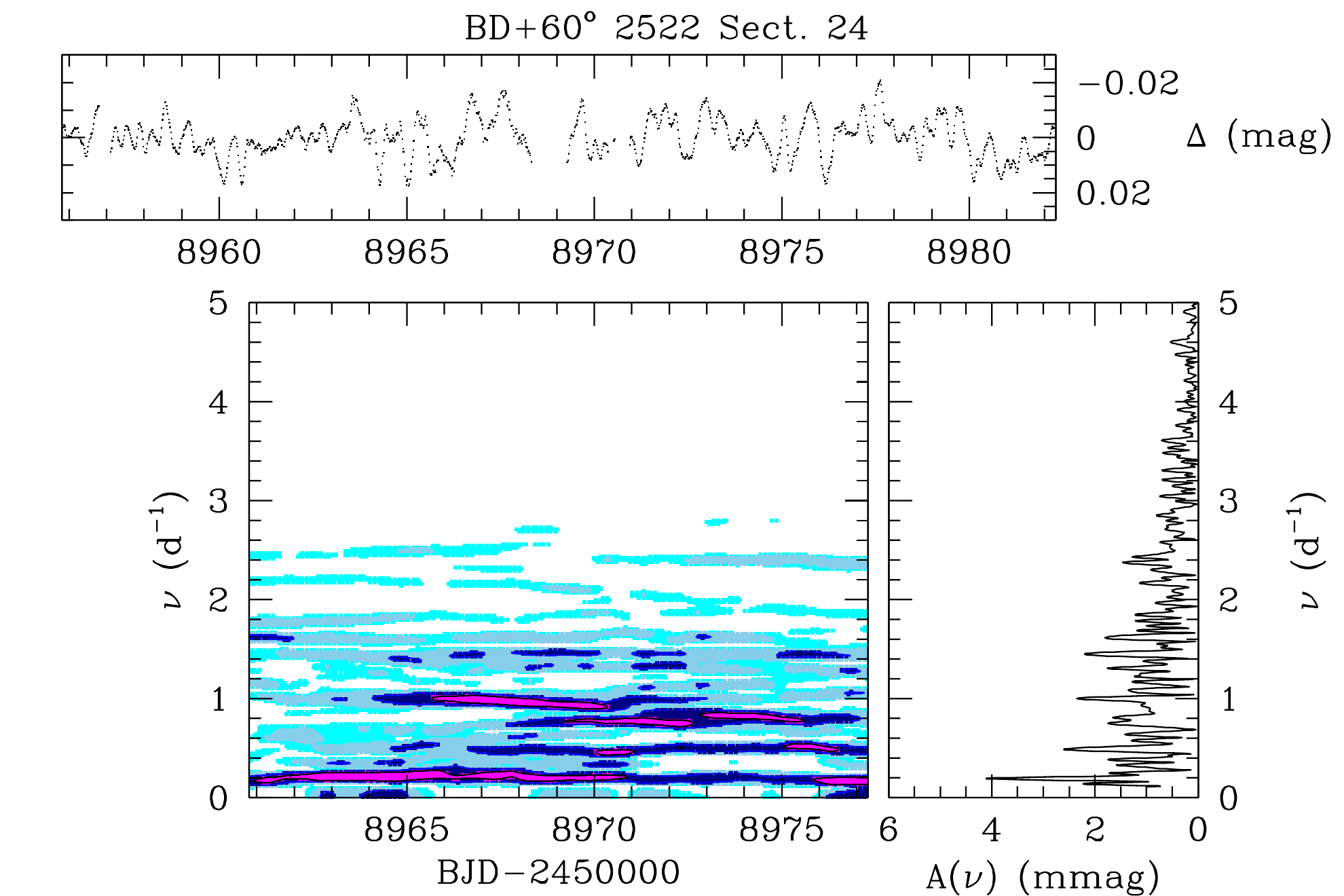}}
  \end{center}
\end{minipage}
\vspace*{5mm}

\begin{minipage}{8.5cm}
  \begin{center}
    \resizebox{!}{5.5cm}{\includegraphics{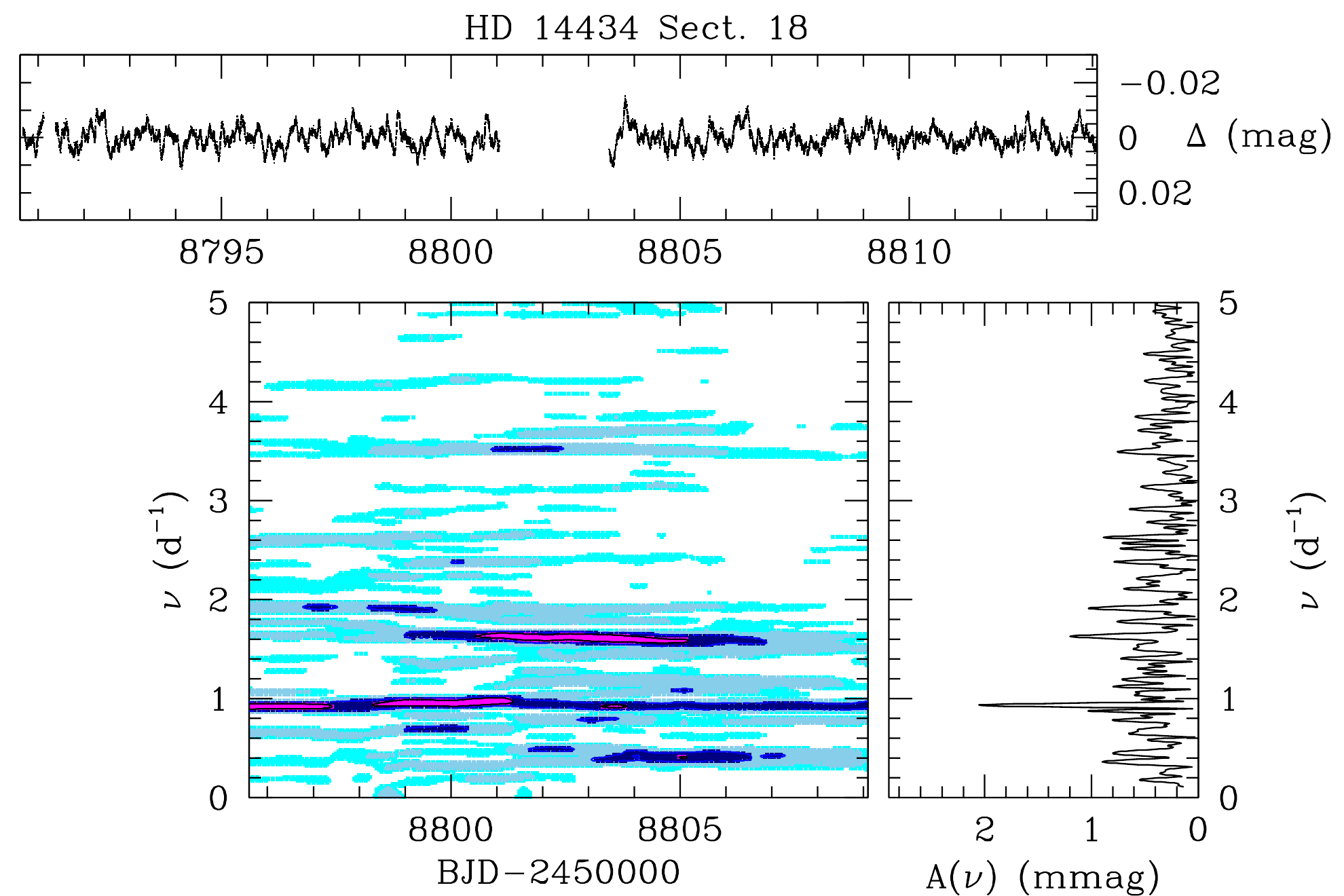}}
  \end{center}
\end{minipage}
\begin{minipage}{8.5cm}
  \begin{center}
    \resizebox{!}{5.5cm}{\includegraphics{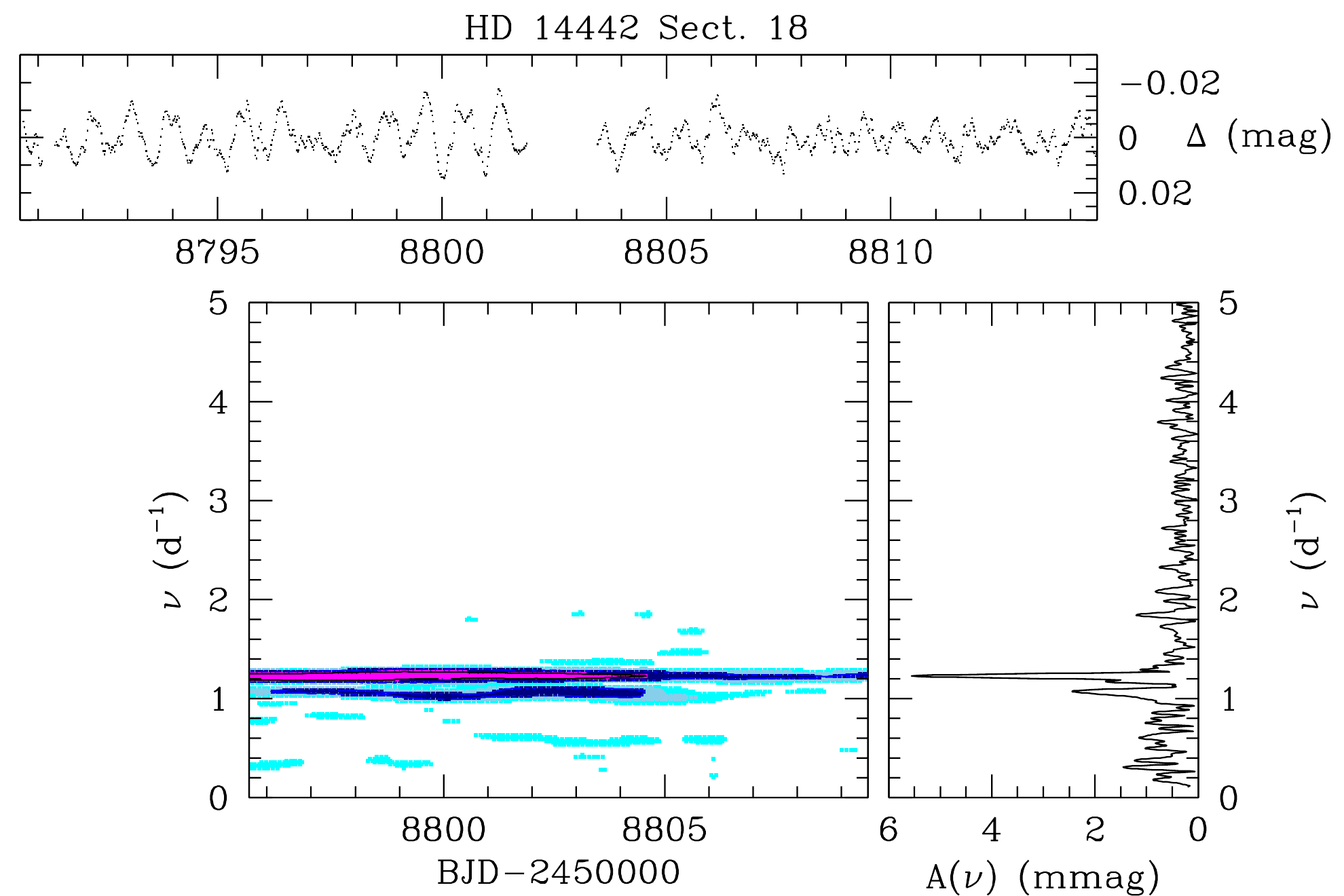}}
  \end{center}
\end{minipage}
\vspace*{5mm}    

\begin{minipage}{8.5cm}
  \begin{center}
    \resizebox{!}{5.5cm}{\includegraphics{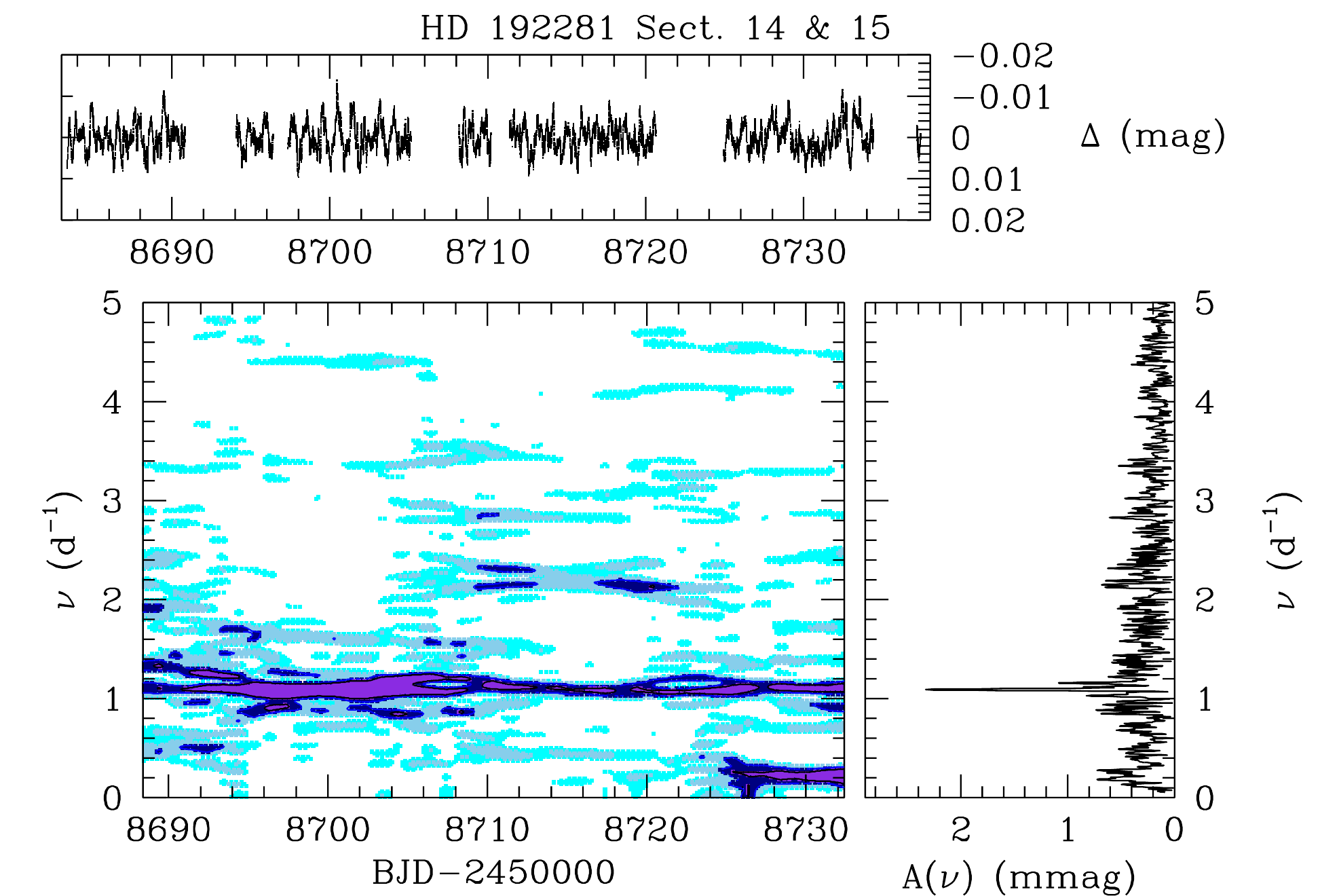}}
  \end{center}
\end{minipage}
\begin{minipage}{8.5cm}
  \begin{center}
    \resizebox{!}{5.5cm}{\includegraphics{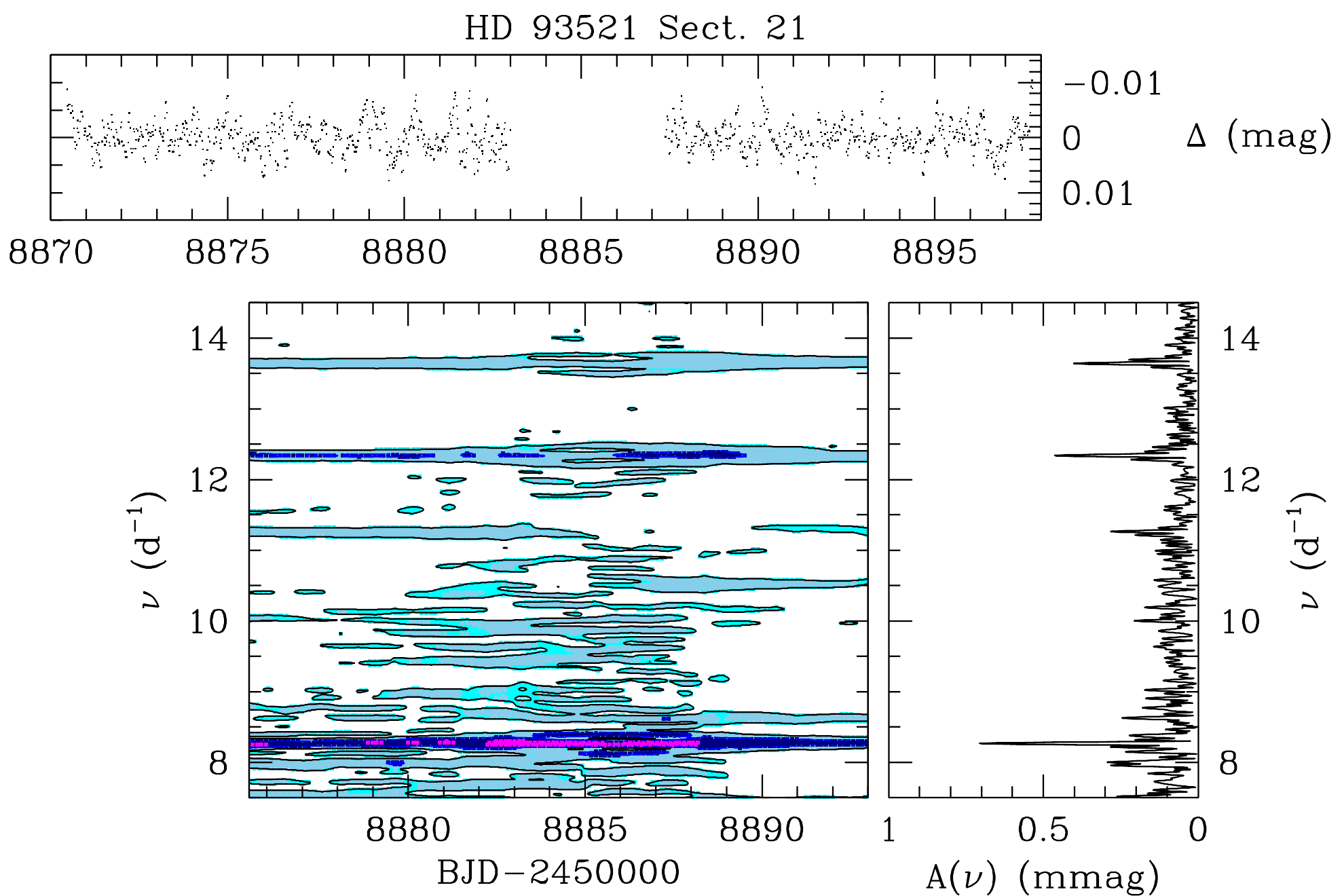}}
  \end{center}
\end{minipage}  
\caption{Time-frequency diagram of the {\it TESS} photometric time series of our target stars. For each time-frequency diagram, the top panel shows the observed lightcurve whilst the bottom left panel provides the evolution of the Fourier periodogram between 0 and 5\,d$^{-1}$ (or between 7.5 and 14.5\,d$^{-1}$ in the case of HD~93\,521) with the date at the middle of the 10-day sliding window. Violet, dark blue, light blue, and cyan stand respectively for amplitudes $\geq 4.0$\,mmag, $\geq 3.0$\,mmag, $\geq 2.0$\,mmag, and $\geq 1.5$\,mmag for BD$+60^{\circ}$\,2522, and amplitudes $\geq 6.0$\,mmag, $\geq 4.0$\,mmag, $\geq 3.0$\,mmag, and $\geq 2.0$\,mmag for HD~14\,434. The same colours indicate amplitudes $\geq 2.0$\,mmag, $\geq 1.5$\,mmag, $\geq 1.0$\,mmag, and $\geq 0.75$\,mmag for HD~14\,434 and HD~192\,281, and amplitudes $\geq 0.75$\,mmag, $\geq 0.5$\,mmag, $\geq 0.25$\,mmag, and $\geq 0.2$\,mmag for HD~93\,521. The right panel of a given time-frequency diagram illustrates the Fourier periodogram evaluated over the full duration of the campaign.\label{FigAppen1}}
\end{figure*}

Below we provide the time-frequency diagrams (Fig.\,\ref{FigAppen1}) for HD~14\,434, HD~14\,442, HD~192\,281, BD$+60^{\circ}$\,2522, and HD~93\,521, computed in the same way as for $\lambda$~Cep (Fig.\,\ref{spevol}).

\section{Unidentified lines in the spectrum of $\lambda$~Cep} \label{Appendix2}
The mean spectrum of $\lambda$~Cep shown in Fig.\,\ref{meanspec} reveals the presence of a weak absorption feature on the blue side of the He\,{\sc i} $\lambda$\,4713 line. This line was previously noted by \citet{deJ99}, but no identification was proposed. 

Given its wavelength ($\sim 4705.4$\,\AA), the most likely identification in the NIST Atomic Spectra Database\footnote{https://www.nist.gov/pml/atomic-spectra-database} appears to be O\,{\sc ii}. Such an identification could further explain the weak absorptions seen near 4593.1 (which could be a blend of O\,{\sc ii} $\lambda\lambda$\,4591, 4596), 4603.3 and 4609.4\,\AA\ (see Fig.\,\ref{meanspec}). The 4603\,\AA\ absorption was previously attributed to N\,{\sc v} $\lambda$\,4604 \citep{Uuh14}, but the absence of N\,{\sc v} $\lambda$\,4620 renders this identification dubious. The identification of this line and the other three features (including the 4705\,\AA\ line) as being due to O\,{\sc ii} offers an alternative. However, it must be stressed that several additional lines (O\,{\sc ii} $\lambda\lambda$\,4639-4642, 4649-4651, 4662, 4674-4676) should be present with a similar strength. These lines are not detectable, though this could be a result of the complexity of this spectral region which contains lines from many different ions. Moreover, the presence of O\,{\sc ii} lines would be surprising given the effective temperature of $\lambda$~Cep which was found to be $36 \pm 1$\,kK \citep{Bouret,Mar15,Cazorla}.
For such a hot star, the ionization fraction of O\,{\sc ii} is expected to be very low in the photosphere \citep[][Bouret 2020, private communication]{LanHub03}. Hence, these lines remain currently unidentified. The only way to explain their presence would be to assume a rather extreme gravity darkening effect that would reduce the effective temperature at the stellar equator below about 25\,kK \citep{LanHub07}. Using the formalism of \citet{EspLara}, we find that a star rotating at 95\% of the critical velocity would fulfill this requirement. However, this is a rather extreme value for a slightly evolved massive star.

\bsp
\label{lastpage}
\end{document}